\documentclass[11pt]{article}

\usepackage[utf8]{inputenc}

\usepackage{amstext,amsmath,amssymb,amsfonts,bbm}
\usepackage{hyperref}
\usepackage{authblk}
\usepackage{caption} 
\usepackage{epsfig} 
\usepackage{color} 
\usepackage{graphicx} 
\usepackage{braket} 
\usepackage{slashed} 
\usepackage{dsfont} 
\usepackage{amsthm}  
\usepackage{physics}

\usepackage{cite}

\allowdisplaybreaks[4]

\usepackage{psfrag}
\usepackage{tikz} 
\usetikzlibrary{arrows}
\usetikzlibrary{plotmarks}
\usetikzlibrary{external}
\usetikzlibrary{patterns}
\usepackage{pgfplots}
\pgfplotsset{compat=1.9}
\usetikzlibrary{patterns}
\usetikzlibrary{calc}
\usepackage[compat=1.1.0]{tikz-feynman}
\usetikzlibrary{automata,positioning}

\usepackage{float}  
\usepackage{subfig}
\usepackage[lmargin=50pt,rmargin=60pt,tmargin=60pt,bmargin=65pt]{geometry}
\graphicspath{{figure/}}

\newcommand{\be}{\begin{equation}}
\newcommand{\ee}{\end{equation}} 

\newcommand{\f}{\frac}

\let\a=\alpha \let\b=\beta    \let\d=\delta

\let\G=\Gamma     \let\X=F
         
  \let\eps=\epsilon

\newcommand{\phib}{\bar{\phi}}

\newcommand{\cN}{\mathcal{N}}

\newcommand{\cT}{\mathcal{T}}

\newcommand{\mba}{\mathbf{a}}
\newcommand{\mbb}{\mathbf{b}}
\newcommand{\mbc}{\mathbf{c}}
\newcommand{\mbd}{\mathbf{d}}
\newcommand{\mbe}{\mathbf{e}}
\newcommand{\mbf}{\mathbf{f}}
\newcommand{\mbg}{\mathbf{g}}
\newcommand{\mbh}{\mathbf{h}}
\newcommand{\mbj}{\mathbf{j}}
\newcommand{\mbk}{\mathbf{k}}
\newcommand{\mbm}{\mathbf{m}}

\newcommand{\mbp}{\mathbf{p}}
\newcommand{\mbq}{\mathbf{q}}
\newcommand{\mbl}{\mathbf{l}}

\allowdisplaybreaks[4]

\theoremstyle{remark}

\definecolor{orange}{rgb}{0.88,0.39,0.12} 
\definecolor{rouge}{rgb}{0.8, 0.0, 0.0}
\definecolor{vert}{rgb}{0.4, 0.69, 0.2}
\definecolor{bleu}{rgb}{0.19, 0.55, 0.91}
\definecolor{lavenderpurple}{rgb}{0.59, 0.48, 0.71}

\begin{document}

\title{\bf Sextic tensor model in rank $3$ at next-to-leading order}

\author[1,2]{Sabine Harribey}

\affil[1]{\normalsize \it 
 CPHT, CNRS, Ecole Polytechnique, Institut Polytechnique de Paris, Route de Saclay, \authorcr 91128 PALAISEAU, 
 France \authorcr
\hfill }

\affil[2]{\normalsize\it 
Heidelberg University, Institut f\"ur Theoretische Physik, Philosophenweg 19, 69120 Heidelberg, Germany
 \authorcr \hfill
 
\bigskip
Email: sabine.harribey@polytechnique.edu }

\date{}

\maketitle

\hrule\bigskip

\begin{abstract}
We compute the four-loop beta functions of short and long-range multi-scalar models with general sextic interactions and complex fields. We then specialize the beta functions to a $U(N)^3$ symmetry and study the renormalization group at next-to-leading order in $N$ and small $\epsilon$. In the short-range case, $\epsilon$ is the deviation from the critical dimension while it is the deviation from the critical scaling of the free propagator in the long-range case. This allows us to find the $1/N$ corrections to the rank-3 sextic tensor model of \cite{Benedetti:2019rja}. In the short-range case, we still find a non-trivial real IR stable fixed point, with a diagonalizable stability matrix. All couplings, except for the so-called wheel coupling, have terms of order $\epsilon^0$ at leading and next-to-leading order, which makes this fixed point different from the other melonic fixed points found in quartic models. In the long-range case, the corrections to the fixed point are instead not perturbative in $\epsilon$ and hence unreliable; we thus find no precursor of the large-$N$ fixed point.
\end{abstract}
\bigskip
\hrule\bigskip

\tableofcontents

\section{Introduction}

In recent years, interest for tensor models has remarkably grown. The main reason is that they exhibit a melonic large-$N$ limit \cite{Bonzom:2011zz,RTM,Klebanov:2018fzb} which is both richer than that of vector models \cite{Guida:1998bx,Moshe:2003xn} and simpler than the planar limit of matrix models \cite{'tHooft:1973jz,Brezin:1977sv,DiFrancesco:1993nw}. Even though, as algebraic objects, tensors are more complicated than matrices, their large-$N$ limit is simpler as melonic diagrams are a subset of planar diagrams. 

Tensor models were first introduced in zero dimension in the context of quantum gravity and random geometry \cite{Ambjorn:1990ge,Sasakura:1990fs,Gurau:2009tw,Gurau:2010ba,Gurau:2011xq}. They were then used as an analytic tool to explore strong coupling effects in many-body quantum mechanics \cite{Witten:2016iux,Gurau:2016lzk,Klebanov:2016xxf,Peng:2016mxj,Krishnan:2016bvg,Krishnan:2017lra,Bulycheva:2017ilt,Choudhury:2017tax,Halmagyi:2017leq,Klebanov:2018nfp,Carrozza:2018psc} (see also \cite{Delporte:2018iyf,Klebanov:2018fzb} for reviews). In one dimension, for example, they were useful to provide an alternative to the SYK model without the quenched disorder of the latter \cite{Sachdev:1992fk, Kitaev2015, Maldacena:2016hyu, Polchinski:2016xgd,Gross:2016kjj}. Finally, tensor models were generalized in higher dimensions where they can be studied as proper field theories. An interesting feature is that they give rise to a new class of conformal field theories, called melonic CFTs \cite{Giombi:2017dtl,Prakash:2017hwq,Benedetti:2017fmp,Giombi:2018qgp,Benedetti:2018ghn,Benedetti:2019eyl,Benedetti:2019ikb,Benedetti:2019rja,Lettera:2020uay} (see also \cite{Benedetti:2020seh,Gurau:2019qag} for reviews). 

This was first observed in dimensional regularization for a short-range model in \cite{Giombi:2017dtl}. The existence of a melonic fixed point was then proven for a long-range quartic $O(N)^3$ model in \cite{Benedetti:2019eyl}. Conformal properties and next-to-leading order corrections were then computed in \cite{Benedetti:2019ikb} and in \cite{Benedetti:2020sye}. However, there have been less studies for sextic models. They were first considered without studying the fixed points and in rank $5$ in \cite{Giombi:2017dtl}. In \cite{Giombi:2018qgp}, a $O(N)^3$ bosonic tensor model was considered and a so-called "prismatic" fixed point was found in $d=3-\epsilon$ dimensions. $1/N$ corrections to this real stable fixed point were also computed. Two models (in rank 3 and 5) were then studied at large $N$ in \cite{Benedetti:2019rja} with the optimal scaling defined in \cite{Carrozza:2015adg} differently to what was done in \cite{Giombi:2018qgp}. A non-trivial fixed point was found for the sextic model in rank $3$ in both the short and the long-range settings. However, in rank $5$, only a non-perturbative fixed point was found. 

In this paper, we go one step further and compute the next-to-leading order contributions to the sextic fixed points in rank $3$ both in short and long range. 

To do so, we start by computing the beta functions of a generic sextic multi-scalar model with $\mathcal{N}$ complex fields. Sextic vector models with a $O(\mathcal{N})$ symmetry were studied both at large and finite $\mathcal{N}$ \cite{Bardeen:1983rv,Amit:1984ri,Pisarski:1982vz,Pisarski:1983gn,Jack:2020wvs}. Computations up to six-loops were carried out in \cite{hager2002six}. Generic multi-scalar models with real fields were also considered in \cite{Codello:2018nbe,ODwyer:2007brp,Osborn:2017ucf} as well as hypercubic models in \cite{BenAliZinati:2021rqc} respectively. Here, we choose complex fields in order to reduce the number of tensor invariants when the symmetry is specified to $U(N)^3$ with $\mathcal{N}=N^3$. After specifying the symmetry, we compute the $1/N$ corrections of the sextic tensor model in rank $3$. 

\paragraph{Main results} In the short-range case, our results are the following. At large $N$ we recover the real interacting fixed point of \cite{Benedetti:2019rja}. At leading-order, the stability matrix has an eigenvalue of multiplicity two and is thus non-diagonalizable: the fixed point corresponds to a logarithmic CFT. At next-to-leading order, the corrections to the fixed point are still real and the stability matrix is now diagonalizable. An interesting feature of this fixed point is that, except for the so-called wheel coupling, all critical couplings have terms of order $\mathcal{O}(\epsilon^0)$. This is thus a different type of fixed point than those found in quartic tensor models. In the long-range case, we again recover the results of \cite{Benedetti:2019rja} at large $N$. However, at NLO, the fixed points are non-perturbative and do not correspond to a precursor of the large-$N$ line of fixed points.

\paragraph{Plan of the paper}

In section \ref{sec:sexticMS}, we study a general short-range sextic multi-scalar model. After defining the action, we compute the wave function renormalization and the beta functions. Then, we apply those results to the $U(N)^3$ sextic tensor model in subsection \ref{sec:tensor_SR}. 

In section \ref{sec:MS_LR}, we study a general long-range sextic multi-scalar model. We again compute the beta functions and finally apply those results to a $U(N)^3$ long-range tensor model in \ref{sec:tensor_LR}.

Finally, we give more details on our computations and conventions in the appendices. In App. \ref{ap:renor}, we detail the computation of the renormalized series. In App. \ref{ap:melon} and App. \ref{ap:betafun4}, we give detailed computations of the melon integrals and of the Feynman integrals appearing in the beta functions. In App. \ref{ap:conventions}, we give the explicit forms of the interaction terms appearing in the action of the short and long-range sextic tensor models. In App. \ref{ap:stability}, we give the full stability matrix for the $U(N)^3$ short-range model. Lastly, in App. \ref{ap:O(N)} we compare our results to previous results obtained for the short-range sextic $O(N)$ model.

\section{The short-range sextic multi-scalar model}
\label{sec:sexticMS}

\subsection{Action}

The short-range multi-scalar model with sextic interactions and complex fields in dimension $d$ is defined by the action:

\begin{align}
		S[\phi]  \, &= \, \int d^dx \, \bigg[ \frac{1}{2} \partial_{\mu} \bar{\phi}_\mba(x) \partial_{\mu} \phi_{\mba}(x)
		\, + \, \frac{1}{(3!)^2} \, \lambda_{\mba \mbb \mbc ; \mbd \mbe \mbf } \phi_{\mba}(x)\phi_{\mbb}(x)\phi_{\mbc}(x) \bar{\phi}_{\mbd}(x)  \bar{\phi}_{\mbe}(x)  \bar{\phi}_{\mbf}(x) \bigg] \, ,
	\end{align}
where the indices take values from 1 to $\cN$, and a summation over repeated indices is implicit. The coupling $\lambda_{\mba \mbb \mbc ; \mbd \mbe \mbf }$ is fully symmetric into the first three indices (corresponding to fields $\phi$) and the last three indices (corresponding to fields $\bar{\phi}$). It thus amounts in general to $2\binom{\cN+2}{3}=\frac{\cN(\cN+1)(\cN+2)}{3}$ couplings. 

In $d=3$, the sextic interaction is marginal. Two and four-point functions are power divergent while six-point functions are logarithmically divergent in the UV. Correlation functions with more than eight external points are trivially convergent. 

In the following, we will work in dimensional regularization with $d=3-\epsilon$ and are interested in Wilson-Fisher like fixed points with $\epsilon\neq 0$ but small. This allows UV divergences to be regularized. Moreover, we choose the BPHZ zero momentum subtraction scheme. One could wonder why we chose this prescription and not the usual Gell-Mann and Low subtraction at non-zero momentum. The reason is that, with this prescription, we were not able to obtain analytic results for the amplitudes of four-loop diagrams in the long-range case. For consistency, we chose the zero momentum subtraction scheme for both short and long-range models.

However, as we work with a massless propagator, we also need an IR regulator. Following \cite{Benedetti:2020rrq}, we introduce an IR regulator by modifying the covariance as\footnote{We denote $x,y$ and so on positions, and $\int_x \equiv \int d^d x$. We denote $p,q$ and so on momenta and $\int_p \equiv \int \frac{d^d p}{(2\pi)^d}$. The Fourier transform is $f(p)=\int_x e^{i px}f(x)$ with inverse $f(x)=\int_p e^{-i px}f(p)$.}:

\begin{equation}
C_{\mu}(p)=\frac{1}{p^2+\mu^2}=\int_0^{\infty} da \,e^{-ap^2-a\mu^2}
\label{eq:free_cov}
\end{equation}
for some mass parameter $\mu>0$.
We give the modified covariance in momentum space as in the following the beta functions will be computed from the momentum space Lagragian. We also explicitly wrote the integral representation of the covariance with Schwinger parameters as we will use it to compute Feynman amplitudes.

Working in dimensional regularization we can renormalize the mass terms to zero. Moreover, there are no divergences in the four-point kernel (see \cite{Benedetti:2019rja} for a detailed proof). We can thus take the four-point couplings to be zero from the beginning.  

\subsection{Wave function renormalization}

We introduce the wave function renormalization by rescaling the bare field $\phi_{\mba}$ as $\phi_{\mba}=\left(\sqrt{Z}\right)_{\mba \mbb}\phi^R_{\mbb}$ with $\phi^R_{\mbb}$ the renormalized field.

\begin{figure}[htbp]
\centering
\includegraphics[scale=1]{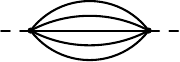}
\caption{Two-point melon graph contributing to the wave function renormalization.}
\label{fig:melonSDE}
\end{figure}

In dimensional regularization, $d=3-\epsilon$, there is only one Feynman graph contributing at lowest order: the melon graph (see Fig. \ref{fig:melonSDE}). 

We then have for the expansion of the renormalized two-point function at lowest order:
\begin{equation}
\Gamma^{(2)}_{\mba \mbb}(p)\equiv G^R_{\mba \mbb}(p)^{-1} =Z_{\mba \mbb}p^2-\frac{\lambda_{\mba \mbc \mbd ; \mbe \mbf \mbg}\lambda_{\mbe \mbf \mbg ; \mbc \mbd \mbb}}{24} M_1(p) \,.
\label{eq:gamma2}
\end{equation}
with $M_1(p)$ the melonic integral. It is computed in App.~\ref{ap:melon}, Eq.~\eqref{eq:M-1}. At leading order in $\epsilon$, we have:
\begin{equation}
M_1(p)=-p^{2-4\epsilon}\frac{2\pi^2}{3\epsilon(4\pi)^{6}} + \mathcal{O}(\epsilon^0)\, ,
\end{equation}
At last, we fix $Z_{\mba \mbb}$ such that
\be
\lim_{\epsilon\to 0}\frac{d\Gamma^{(2)}_{\mba \mbb}(p)}{dp^2}|_{p^2=\mu^2}=\delta_{\mba \mbb}\,.
\ee

At quadratic order in the coupling constant, we obtain:
\begin{equation}
Z_{\mba \mbb}=\delta_{\mba \mbb}+\frac{\lambda_{\mba \mbc \mbd ;\mbe \mbf \mbg}\lambda_{\mbe \mbf \mbg ; \mbc \mbd \mbb}}{24}\tilde{M}_1(\mu) =\delta_{\mba \mbb}-\mu^{-4\epsilon}\frac{\lambda_{\mba \mbc \mbd ;\mbe \mbf \mbg}\lambda_{\mbe \mbf \mbg ;\mbc \mbd \mbb}\pi^2}{36\epsilon(4\pi)^{6}}\,,
\label{eq:wavef3}
\end{equation}
with $\tilde{M}_1(\mu)=\frac{d}{dp^2}M_1(p)|_{p^2=\mu^2}$.

\subsection{Beta functions}

We define the renormalized sextic coupling $g_{\mba \mbb \mbc ; \mbd \mbe \mbf}$ in terms of the bare expansion of the six-point function by the following renormalization condition:

\begin{equation}
g_{\mba \mbb \mbc ;\mbd \mbe \mbf}=\mu^{-2\epsilon}\Gamma^{(6)}_{\mbg \mbh \mbj ;\mbk \mbp \mbq}(0,\dots,0)\left(\sqrt{Z}\right)_{\mbg \mba}\left(\sqrt{Z}\right)_{\mbh \mbb}\left(\sqrt{Z}\right)_{\mbj \mbc}\left(\sqrt{Z}\right)_{\mbk \mbd} \left(\sqrt{Z}\right)_{\mbp \mbe}\left(\sqrt{Z}\right)_{\mbq \mbf} \, ,
\end{equation}
where $\Gamma^{(6)}_{\mbg \mbh \mbj ;\mbk \mbp \mbq}(0,\dots,0)$ is the one-particle irreducible six-point function at zero external momentum. We compute it up to order three in the coupling constant (four-loops) using the bare expansion in terms of connected amputated one particle irreducible Feynman graphs. We write the amplitude of the latter in Schwinger parametrization as \cite{Benedetti:2020rrq}:

\be\label{eq:amp_final}
\mathcal{A}(G ) =  \mu^{ (2d-6\zeta)(V-1)} \; \mathcal{\hat{A}}(G) \,, \quad
\mathcal{\hat{A}}(G) =
 \frac{1} { 
  \big[ (4\pi)^{d} \Gamma(\zeta)^3 \big]^{V-1} }
\int_0^{\infty}
\prod_{e \in G} d a_e
\;\;
\frac{\prod_{e \in G} a_e^{\zeta-1} \; e^{-\sum_{e \in G} a_e}}
{\big(\sum_{\cT \in G  } \prod_{e \notin \cT } a_e\big)^{d/2} } \,,
\ee
where $V$ denote the numbers of vertices of $G$, $e \in G$ runs over the edges of $G$, and $\cT$ denotes the spanning trees in $G$ (e.g.\ \cite{Rivasseau:1991ub}). Note that we used the fact that we only deal with six-point graphs with sextic vertices, as these are sufficient to describe the divergent graphs described above. The dependence in $\mu$ was found by rescaling all internal momenta by $\mu$ before introducing the Schwinger parametrization.

\begin{figure}[htbp]
\centering
\captionsetup[subfigure]{labelformat=empty}
\subfloat[$D_{\zeta}$]{\includegraphics[scale=1]{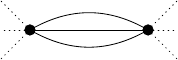}}
\hspace{1cm}
\subfloat[$D_{\zeta}^2$]{\includegraphics[scale=1]{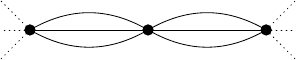}}
\\
\subfloat[$S_{\zeta}$]{\includegraphics[scale=1]{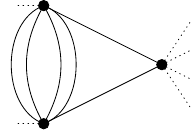}}
\hspace{1cm}
\subfloat[$I_{\zeta}$]{\includegraphics[scale=1]{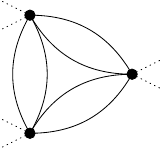}}
\hspace{1cm}
\subfloat[$J_{\zeta}$]{\includegraphics[scale=1]{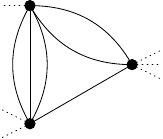}}
\caption{The five graphs contributing to the bare expansion up to order $3$ in the coupling constant. For the short-range case, $\zeta=1$.}
\label{fig:graphs}
\end{figure}

There are five graphs contributing to the bare expansion up to order $3$ in the coupling constant. They are represented in Fig.~\ref{fig:graphs} and we call $D_{\zeta},S_{\zeta},I_{\zeta},J_{\zeta}$ the amplitudes $\mathcal{\hat{A}}(G)$ of these diagrams. These are the amplitudes without the $\mu$ dependence that has been factored out. For the second diagram in Fig. \ref{fig:graphs}, we use the fact that the amplitude of a one-vertex reducible diagram factors into the product of the amplitudes of its one-vertex irreducible parts.

We should also be careful to conserve the permutation symmetry of the six-point function in its indices. Therefore, we should completely symmetrize over external white and black vertices. However, due to specific symmetries of the diagrams under relabeling, certain terms are equal. Grouping them together and setting $\zeta=1$, we get:

\begin{align}
&\Gamma^{(6)}_{\mba \mbb \mbc ;\mbd \mbe \mbf}(p_1,\dots,p_6)=\lambda_{\mba \mbb \mbc ;\mbd \mbe \mbf}-\frac{1}{6}\left[ 3\left(\lambda_{\mba \mbb \mbg ;\mbh \mbj \mbd}\lambda_{\mbc \mbh \mbj ;\mbe \mbf \mbg}+ 8\text{ terms}\right) +\lambda_{\mba \mbb \mbc ;\mbg \mbh \mbj}\lambda_{\mbg \mbh \mbj ;\mbd \mbe \mbf}\right]\mu^{-2\epsilon}D_1 \crcr
& +\frac{1}{12}\left[3\left(\lambda_{\mba \mbg \mbh ; \mbj \mbk \mbl}\lambda_{\mbb \mbj \mbk ;\mbg \mbh \mbm}\lambda_{\mbl \mbm \mbc ;\mbd \mbe \mbf} + 2\text{ terms}\right)+ 3\left(\lambda_{ \mbj \mbk \mbl ; \mbd \mbg \mbh}\lambda_{ \mbg \mbh \mbm ;\mbe \mbj \mbk}\lambda_{ \mba \mbb \mbc ;\mbf \mbl \mbm} + 2\text{ terms}\right) \right.\crcr
&\left.  \qquad + 3\left(\lambda_{\mba \mbg \mbh ;\mbj \mbk \mbl}\lambda_{ \mbj \mbk \mbm ;\mbg \mbh \mbd}\lambda_{\mbl \mbb \mbc ;\mbe \mbf \mbm} + 8\text{ terms}\right)+2\left(\lambda_{\mba \mbg \mbh ;\mbj \mbk \mbl }\lambda_{ \mbj \mbk \mbl ;\mbh \mbm \mbd}\lambda_{\mbm \mbb \mbc ;\mbe \mbf \mbg} + 8\text{ terms}\right)\right]\mu^{-4\epsilon}S_1 \crcr
& +\frac{1}{36}\left[ 9\left(\lambda_{\mba \mbg \mbh ;\mbj \mbe \mbf}\lambda_{\mbj \mbk \mbl ;\mbm \mbg \mbh}\lambda_{\mbm \mbb \mbc ;\mbk \mbl \mbd}+ 8\text{ terms}\right) +\lambda_{\mba \mbb \mbc ;\mbg \mbh \mbj}\lambda_{\mbg \mbh \mbj ;\mbk \mbl \mbm}\lambda_{\mbk \mbl \mbm ;\mbd \mbe \mbf}\right]\mu^{-4\epsilon}D_1^2 \crcr
& +\frac{1}{4}\left[\left(\lambda_{\mba \mbg \mbh ; \mbd \mbj \mbk}\lambda_{\mbb \mbc \mbl ;\mbg  \mbh \mbm}\lambda_{\mbj  \mbk \mbm ; \mbl \mbe \mbf} + 8 \text{ terms}\right)+ 2\left(\lambda_{\mba \mbg  \mbh ;\mbd  \mbj \mbk }\lambda_{\mbb \mbc \mbj; \mbg \mbl \mbm}\lambda_{\mbk  \mbl \mbm ;\mbh \mbe \mbf} + 8 \text{ terms}\right) \right. \crcr
& \left. \qquad +\left(\lambda_{\mba \mbg \mbh ;\mbd  \mbj \mbk}\lambda_{\mbb \mbl \mbm ;\mbe \mbg \mbh}\lambda_{\mbc  \mbj \mbk ;\mbf \mbl \mbm} + 5 \text{ terms}\right)+4\left(\lambda_{\mba \mbg \mbh ; \mbd  \mbj \mbk}\lambda_{\mbb \mbj \mbl ;\mbe \mbg \mbm}\lambda_{\mbc  \mbk \mbm ;\mbf \mbh \mbl} + 5 \text{ terms}\right)\right]\mu^{-4\epsilon}I_1 \crcr
& +\frac{1}{12}\left[\left(\lambda_{\mba \mbg \mbh ;\mbj \mbe \mbf }\lambda_{\mbb \mbc \mbj ;\mbk \mbl \mbm }\lambda_{\mbk \mbl \mbm ; \mbg \mbh   \mbd} +  \text{ 8 terms}\right) + 6\left(\lambda_{\mba \mbg \mbh ;\mbj \mbe \mbf }\lambda_{\mbb \mbc \mbk ;\mbg \mbl \mbm }\lambda_{\mbl \mbm\mbj ;\mbh \mbk  \mbd} +  \text{ 8 terms}\right)\right. \crcr
& \qquad+ \left(\lambda_{\mbj \mbb \mbc ;\mbd \mbg \mbh }\lambda_{\mbk \mbl \mbm ;\mbe \mbf \mbj  }\lambda_{\mba \mbg \mbh ;\mbk \mbl \mbm} +  \text{ 8 terms}\right) + 6\left(\lambda_{\mbj \mbb \mbc ;\mbd  \mbg \mbh }\lambda_{\mbg \mbl \mbm ;\mbk \mbe \mbf }\lambda_{\mba \mbh \mbk ;\mbl  \mbm \mbj} +  \text{ 8 terms}\right)\crcr
& \qquad +3\left(\lambda_{\mba \mbk \mbl ;\mbm \mbe \mbd}\lambda_{\mbb \mbj \mbm  ;\mbf \mbg  \mbh }\lambda_{\mbc \mbg \mbh ; \mbj  \mbk \mbl} +  \text{ 17 terms}\right)+6\left(\lambda_{\mba \mbl \mbm ;\mbh \mbd \mbe }\lambda_{\mbb  \mbj \mbk ;\mbg\mbm  \mbf }\lambda_{\mbc \mbg \mbh ;\mbj  \mbk \mbl} +  \text{ 17 terms}\right)\crcr
& \qquad +3\left(\lambda_{ \mbm \mba \mbb ;\mbd \mbk \mbl}\lambda_{\mbc   \mbg \mbh ;\mbe \mbj  \mbm}\lambda_{\mbj \mbk \mbl  ;\mbg \mbh   \mbf} +  \text{ 17 terms}\right)+6\left(\lambda_{ \mba \mbb  \mbh ;\mbd \mbl\mbm}\lambda_{\mbc   \mbg \mbm ;\mbe \mbj   \mbk}\lambda_{\mbj \mbk \mbl ;\mbg \mbh   \mbf} +  \text{ 17 terms}\right) \crcr
& \qquad \left. +3\left(\lambda_{\mba \mbb \mbc ;\mbg \mbh \mbj}\lambda_{\mbj \mbl \mbm ;\mbk \mbd \mbe}\lambda_{\mbg \mbh \mbk ;\mbl \mbm \mbf} + \text{ 2 terms}\right)+ 3\left(\lambda_{\mbg \mbh \mbj ;\mbd \mbe \mbf}\lambda_{ \mba \mbb \mbk ;\mbj \mbl \mbm}\lambda_{ \mbl \mbm \mbc ;\mbg \mbh \mbk }+ \text{ 2 terms}\right)\right]\mu^{-4\epsilon}J_1 \, ,
\label{eq:bare_series}
\end{align}
where the notation $+ \dots$ terms corresponds to a sum over terms obtained by permuting external indices into non-equivalent ways. For example, the nine terms in the first line correspond to the choice of the white index $\mba, \mbb$ or $\mbc$ on the second coupling and  the choice of the black index $\mbd,\mbe$ or $\mbf$ on the first coupling.
The integrals $D_{\zeta},S_{\zeta},I_{\zeta},J_{\zeta}$ are computed in App.~\ref{ap:betafun4} both for the short-range case $\zeta=1$ and the long-range case $0<\zeta<1$.

The beta function is then the scale derivative of the running coupling at fixed bare coupling:
\begin{equation}
\beta_{\mba \mbb \mbc ; \mbd \mbe \mbf}=\mu\partial_{\mu}g_{\mba \mbb \mbc ;\mbd \mbe \mbf} \, .
\end{equation}

We rescale the couplings by $\tilde{g}_{\mba \mbb \mbc ;\mbd \mbe \mbf}=(4\pi)^{-d}g_{\mba \mbb \mbc ;\mbd \mbe \mbf}$ and we define:

\begin{align}
\alpha_{D_1}&=\epsilon(4\pi)^d\frac{D_1}{3}=\frac{2\pi}{3} \; ,  \; \alpha_{S_1}=-\epsilon(4\pi)^{2d} \frac{S_1}{3}= \frac{2\pi^2}{3} \; ,  \; \alpha_{I_1}=-\epsilon(4\pi)^{2d}I_1= -\pi^4 \; , \crcr
\alpha_{J_1}&=\epsilon(4\pi)^{2d}\frac{(D_1^2-2J_1)}{6}= -\frac{4\pi^2}{3}  \; , \; \alpha_{M_1}=-\epsilon(4\pi)^{2d}\frac{\tilde{M}_1}{12}=\frac{\pi^2}{18} \, .
\end{align}

We finally obtain the following beta function:
\begin{align}
\beta_{\mba \mbb \mbc ;\mbd \mbe \mbf}&=-2\epsilon \tilde{g}_{\mba \mbb \mbc ;\mbd \mbe \mbf}+\alpha_{D_1}\left[ 3\left(\tilde{g}_{\mba \mbb \mbg ;\mbh \mbj \mbd}\tilde{g}_{\mbc \mbh \mbj ;\mbe \mbf \mbg}+ 8\text{ terms}\right) +\tilde{g}_{\mba \mbb \mbc ;\mbg \mbh \mbj}\tilde{g}_{\mbg \mbh \mbj ;\mbd \mbe \mbf}\right] \crcr
& +\alpha_{S_1}\left[3\left(\tilde{g}_{\mba \mbg \mbh ; \mbj \mbk \mbl}\tilde{g}_{\mbb \mbj \mbk ;\mbg \mbh \mbm}\tilde{g}_{\mbl \mbm \mbc ;\mbd \mbe \mbf} + 2\text{ terms}\right)+ 3\left(\tilde{g}_{ \mbj \mbk \mbl ; \mbd \mbg \mbh}\tilde{g}_{ \mbg \mbh \mbm ;\mbe \mbj \mbk}\tilde{g}_{ \mba \mbb \mbc ;\mbf \mbl \mbm} + 2\text{ terms}\right) \right.\crcr
&\left.  \qquad + 3\left(\tilde{g}_{\mba \mbg \mbh ;\mbj \mbk \mbl}\tilde{g}_{ \mbj \mbk \mbm ;\mbg \mbh \mbd}\tilde{g}_{\mbl \mbb \mbc ;\mbe \mbf \mbm} + 8\text{ terms}\right)+2\left(\tilde{g}_{\mba \mbg \mbh ;\mbj \mbk \mbl }\tilde{g}_{ \mbj \mbk \mbl ;\mbh \mbm \mbd}\tilde{g}_{\mbm \mbb \mbc ;\mbe \mbf \mbg} + 8\text{ terms}\right)\right] \crcr
& +\alpha_{I_1}\left[\left(\tilde{g}_{\mba \mbg \mbh ; \mbd \mbj \mbk}\tilde{g}_{\mbb \mbc \mbl ;\mbg  \mbh \mbm}\tilde{g}_{\mbj  \mbk \mbm ; \mbl \mbe \mbf} + 8 \text{ terms}\right)+ 2\left(\tilde{g}_{\mba \mbg  \mbh ;\mbd  \mbj \mbk }\tilde{g}_{\mbb \mbc \mbj; \mbg \mbl \mbm}\tilde{g}_{\mbk  \mbl \mbm ;\mbh \mbe \mbf} + 8 \text{ terms}\right) \right. \crcr
& \left. \qquad +\left(\tilde{g}_{\mba \mbg \mbh ;\mbd  \mbj \mbk}\tilde{g}_{\mbb \mbl \mbm ;\mbe \mbg \mbh}\tilde{g}_{\mbc  \mbj \mbk ;\mbf \mbl \mbm} + 5 \text{ terms}\right)+4\left(\tilde{g}_{\mba \mbg \mbh ; \mbd  \mbj \mbk}\tilde{g}_{\mbb \mbj \mbl ;\mbe \mbg \mbm}\tilde{g}_{\mbc  \mbk \mbm ;\mbf \mbh \mbl} + 5 \text{ terms}\right)\right]\crcr
& +\alpha_{J_1}\left[\left(\tilde{g}_{\mba \mbg \mbh ;\mbj \mbe \mbf }\tilde{g}_{\mbb \mbc \mbj ;\mbk \mbl \mbm }\tilde{g}_{\mbk \mbl \mbm ; \mbg \mbh   \mbd} +  \text{ 8 terms}\right) + 6\left(\tilde{g}_{\mba \mbg \mbh ;\mbj \mbe \mbf }\tilde{g}_{\mbb \mbc \mbk ;\mbg \mbl \mbm }\tilde{g}_{\mbl \mbm\mbj ;\mbh \mbk  \mbd} +  \text{ 8 terms}\right)\right. \crcr
& \qquad \left(\tilde{g}_{\mbj \mbb \mbc ;\mbd \mbg \mbh }\tilde{g}_{\mbk \mbl \mbm ;\mbe \mbf \mbj  }\tilde{g}_{\mba \mbg \mbh ;\mbk \mbl \mbm} +  \text{ 8 terms}\right) + 6\left(\tilde{g}_{\mbj \mbb \mbc ;\mbd  \mbg \mbh }\tilde{g}_{\mbg \mbl \mbm ;\mbk \mbe \mbf }\tilde{g}_{\mba \mbh \mbk ;\mbl  \mbm \mbj} +  \text{ 8 terms}\right)\crcr
& \qquad +3\left(\tilde{g}_{\mba \mbk \mbl ;\mbm \mbe \mbd}\tilde{g}_{\mbb \mbj \mbm  ;\mbf \mbg  \mbh }\tilde{g}_{\mbc \mbg \mbh ; \mbj  \mbk \mbl} +  \text{ 17 terms}\right)+6\left(\tilde{g}_{\mba \mbl \mbm ;\mbh \mbd \mbe }\tilde{g}_{\mbb  \mbj \mbk ;\mbg\mbm  \mbf }\tilde{g}_{\mbc \mbg \mbh ;\mbj  \mbk \mbl} +  \text{ 17 terms}\right)\crcr
& \qquad +3\left(\tilde{g}_{ \mbm \mba \mbb ;\mbd \mbk \mbl}\tilde{g}_{\mbc   \mbg \mbh ;\mbe \mbj  \mbm}\tilde{g}_{\mbj \mbk \mbl  ;\mbg \mbh   \mbf} +  \text{ 17 terms}\right)+6\left(\tilde{g}_{ \mba \mbb  \mbh ;\mbd \mbl\mbm}\tilde{g}_{\mbc   \mbg \mbm ;\mbe \mbj   \mbk}\tilde{g}_{\mbj \mbk \mbl ;\mbg \mbh   \mbf} +  \text{ 17 terms}\right) \crcr
& \qquad \left. +3\left(\tilde{g}_{\mba \mbb \mbc ;\mbg \mbh \mbj}\tilde{g}_{\mbj \mbl \mbm ;\mbk \mbd \mbe}\tilde{g}_{\mbg \mbh \mbk ;\mbl \mbm \mbf} + \text{ 2 terms}\right)+ 3\left(\tilde{g}_{\mbg \mbh \mbj ;\mbd \mbe \mbf}\tilde{g}_{ \mba \mbb \mbk ;\mbj \mbl \mbm}\tilde{g}_{ \mbl \mbm \mbc ;\mbg \mbh \mbk }+ \text{ 2 terms}\right)\right]\crcr
&+\alpha_{M_1}\left(\tilde{g}_{\mba \mbg \mbh ;\mbj \mbk \mbl}\tilde{g}_{\mbj \mbk \mbl ;\mbg \mbh \mbm}\tilde{g}_{\mbm \mbb \mbc; \mbd \mbe \mbf} + 5 \text{ terms}\right)\, .
\label{eq:beta_SR}
\end{align}

We also compute the field critical exponent defined by:
\begin{equation}
\eta_{\mba \mbb}=2\beta_{\mbk \mbc \mbd ;\mbe \mbf \mbg}\left(\frac{\partial Z^{1/2}}{\partial \tilde{g}_{\mbk \mbc \mbd ;\mbe \mbf \mbg}}Z^{-1/2}\right)_{\mba \mbb}
\end{equation}

Using \eqref{eq:wavef3} and \eqref{eq:beta_SR}, we obtain:
\begin{equation}
\eta_{\mba \mbb}=\frac{\pi^2}{9}\tilde{g}_{\mba \mbc \mbd ;\mbe \mbf \mbg}\tilde{g}_{\mbe \mbf \mbg; \mbc \mbd \mbb}
\end{equation}

By imposing various symmetry restrictions on the interaction, one obtains different models. We study here the case with $U(N)^3$ invariance in order to obtain subleading corrections to the fixed point of \cite{Benedetti:2019rja}.

\subsection{Application: $U(N)^3$ symmetry}
\label{sec:tensor_SR}

\begin{figure}[htbp]
\centering
\captionsetup[subfigure]{labelformat=empty}
\subfloat[$\delta^{(1)}_{\mba \mbb  \mbc ;\mbd \mbe\mbf}$]{\includegraphics[scale=0.5]{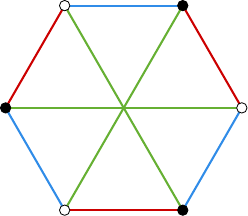}}
\hspace{1cm}
\subfloat[$\delta^{(2)}_{\mba \mbd ; \mbb \mbe; \mbc\mbf}$]{\includegraphics[scale=0.5]{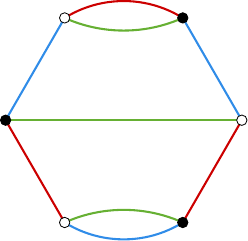}}
\hspace{1cm}
\subfloat[$\delta^{(3)}_{\mba \mbd ; \mbb \mbe; \mbc\mbf}$]{\includegraphics[scale=0.5]{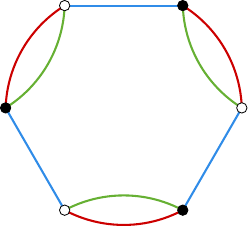}}
\hspace{1cm}
\subfloat[$\delta^{(4)}_{\mba \mbd ; \mbb \mbe; \mbc\mbf}$]{\includegraphics[scale=0.5]{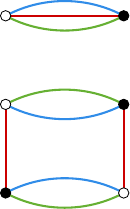}}
\hspace{1cm}
\subfloat[$\delta^{(5)}_{\mba \mbd ; \mbb \mbe; \mbc\mbf}$]{\includegraphics[scale=0.5]{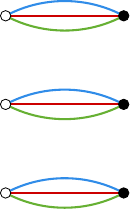}}
\caption{The five $U(N)^3$ invariants. Each white vertex represents a field $\phi$ while each black vertex represents a field $\bar{\phi}$. Each edge corresponds to a contraction of indices and is assigned a color corresponding to the position of the indices in the tensor. The first invariant starting from the left is called the \textit{wheel} invariant.}
\label{fig:invariants}
\end{figure}

In this subsection, we specify the symmetry to $U(N)^3$ with $\mathcal{N}=N^3$. Each index is now a triplet of indices going from $1$ to $N$. There are five different invariants, thus five couplings. We set\footnote{The normalization was chosen so that the couplings are normalized by $1/6$ as usually done in sextic tensor models.}:

\begin{align}
\tilde{g}_{\mba \mbb  \mbc ;\mbd \mbe\mbf}&=\tilde{g}_1 \left(\delta^{(1)}_{\mba \mbb  \mbc ;\mbd \mbe\mbf}+ 5 \text{ terms}\right)+\frac{1}{2}\tilde{g}_2\left( \delta^{(2)}_{\mba \mbd ; \mbb \mbe; \mbc\mbf} + 11 \text{ terms}\right) +\frac{1}{2}\tilde{g}_3 \left(\delta^{(3)}_{\mba \mbd ; \mbb \mbe; \mbc\mbf}+11 \text{ terms}\right)\crcr
& +\tilde{g}_4 \left(\delta^{(4)}_{\mba \mbd ; \mbb \mbe; \mbc\mbf}+ 5 \text{ terms}\right)+\tilde{g}_5 \left(\delta^{(5)}_{\mba \mbd ; \mbb \mbe; \mbc\mbf}+ 5 \text{ terms}\right)\, ,
\label{eq:invariants}
\end{align}
where the contractions are specified in App.~\ref{ap:conventions}. The corresponding invariants are represented in Fig.~\ref{fig:invariants}. The first one is called the \textit{wheel} and we will refer to the associated coupling $\tilde{g}_1$ as the wheel coupling. 

We rescale the couplings as:

\begin{equation}
\tilde{g}_1=\frac{\bar{g}_1}{N^3} \; ,\; \tilde{g}_2=\frac{\bar{g}_2}{N^4}  \; , \; \tilde{g}_3=\frac{\bar{g}_3}{N^4}  \; , \; \tilde{g}_4=\frac{\bar{g}_4}{N^5} \; , \; \tilde{g}_5=\frac{\bar{g}_5}{N^6} \, .
\end{equation}

This rescaling ensures that the model admits a well-behaved large-$N$ expansion (see \cite{Prakash:2019zia} for a rigorous proof).

We then obtain the following beta functions up to order $N^{-1}$:

\begin{align}
\beta_1&=-2 \bar{g}_1\left(\epsilon-\bar{g}_1^2\pi^2\right)-\frac{24\pi^4}{N}\bar{g}_1^3 +\mathcal{O}(N^{-2}) \, ,\crcr
\beta_2&=-2 \bar{g}_2\left(\epsilon-\bar{g}_1^2\pi^2\right)+ 4\pi^2\bar{g}_1^2\left(\frac{9}{2\pi}+9\bar{g}_1+\bar{g}_2\right) \crcr
& +\frac{4}{N}\Bigg[\frac{\pi}{9}\left(81\bar{g}_1^2+36\bar{g}_1\bar{g}_2+\bar{g}_2^2+6\bar{g}_3(9\bar{g}_1+\bar{g}_2)\right)-2\pi^4\bar{g}_1^2\bar{g}_2 -\pi^2\bar{g}_1^2(63\bar{g}_1-2\bar{g}_2+9\bar{g}_3)\Bigg] +\mathcal{O}(N^{-2})\, , \crcr
\beta_3&= -2 \bar{g}_3\left(\epsilon-4\bar{g}_1^2\pi^2\right)+\frac{2}{N}\Bigg[\frac{\pi}{9}\left(36\bar{g}_1\bar{g}_2+2\bar{g}_2^2+9\bar{g}_3^2\right)  +4\pi^2\bar{g}_1^2\left(\bar{g}_2-18\bar{g}_1\right)\Bigg] +\mathcal{O}(N^{-2}) \, ,\crcr
\beta_4&= -2 \bar{g}_4\left(\epsilon-\bar{g}_1^2\pi^2\right)+2\pi^2\bar{g}_1^2\left(27\bar{g}_1+10\bar{g}_2+12\bar{g}_3+7\bar{g}_4\right) \crcr
& + \frac{1}{N}\Bigg[\frac{2\pi}{9}\left(2\bar{g}_2(5\bar{g}_2+12\bar{g}_3+4\bar{g}_4)+36\bar{g}_1(\bar{g}_2+3\bar{g_3}+2\bar{g}_4)+3\bar{g}_3(9\bar{g}_3+4\bar{g}_4)\right) \crcr
& \qquad -\frac{\pi^4}{81}\left(162\bar{g}_1^2(54\bar{g}_1+5\bar{g}_2)+3\bar{g}_3(648\bar{g}_1^2+2\bar{g}_2^2+12\bar{g}_2\bar{g}_3+9\bar{g}_3^2)+2\bar{g}_4(324\bar{g}_1^2+\bar{g}_2^2+18\bar{g}_3^2) \right)\crcr
& \qquad  - 4\pi^2\bar{g}_1^2(36\bar{g}_1+25\bar{g}_2+6\bar{g}_3+6\bar{g}_4) \Bigg] +\mathcal{O}(N^{-2})\, , \crcr
\beta_5&=-2 \bar{g}_5\left(\epsilon-\bar{g}_1^2\pi^2\right)+\pi^2\bar{g}_1^2\left(\frac{2}{\pi}+6\bar{g}_2+16\bar{g}_4+30\bar{g}_5\right) \crcr
& + \frac{1}{N}\Bigg[ -\frac{2\pi^4}{243}\left(3(31\bar{g}_3+7\bar{g}_4)\bar{g}_2^2+10\bar{g}_2^3 +9\bar{g}_3\bar{g}_4(3\bar{g}_3+2\bar{g}_4)+2\bar{g_4}^3+243\bar{g}_1^2(\bar{g}_2+3\bar{g}_3+\bar{g}_4) \right.\crcr
& \qquad \qquad \qquad \left. +6\bar{g}_2(9\bar{g}_3^2+12\bar{g}_3\bar{g}_4+2\bar{g}_4^2)\right) \crcr
& \qquad +\frac{4\pi}{9}\left(\bar{g}_2(\bar{g}_2+6\bar{g}_3+4\bar{g}_4)+2\bar{g}_4(3\bar{g}_3+\bar{g}_4)\right)-4\pi^2\bar{g}_1^2(3\bar{g}_2+3\bar{g}_3+4\bar{g}_4) \Bigg] +\mathcal{O}(N^{-2}) \, .
\end{align}

Similarly, we find for the field critical exponent:
\begin{equation}
\eta=\frac{2\pi^2\bar{g}_1^2}{3}+\mathcal{O}(N^{-2}) \, .
\end{equation}

If we try to solve naively these beta functions, we find non-perturbative fixed points that blow up when we send $\epsilon \rightarrow 0$. For example, $g_2^{\star}$ has the following form:
\begin{equation}
g_2^{\star}=a+b\sqrt{\epsilon}+\frac{c}{\epsilon N}+\mathcal{O}(N^{-2}) \, .
\end{equation}
with $a,b,c$ constants of order $1$. 

It is because here the behavior of the fixed point is governed by the combination $\epsilon N$. Indeed, the fixed points of the typical melonic large-$N$ limit are obtained for $1 \gg \frac{1}{\epsilon N}$ or $\epsilon N \gg 1$. As we wish to study the $1/N$ corrections to these fixed points we set:
\begin{equation}
\tilde{N}=\epsilon N \, ,
\end{equation}
and we expand first in $1/\tilde{N}$ and then in $\epsilon$. This is very similar to what happened in \cite{Benedetti:2020sye}. 

We parametrize the critical couplings as $\bar{g}_i=\bar{g}_{i,0}+\frac{\bar{g}_{i,1}}{\tilde{N}} +\mathcal{O}(\tilde{N}^{-2})$ for $i=1,\dots ,5$. Solving for the zeros of the beta functions at leading order, apart from the Gaussian fixed point, we find the following solutions:
\begin{gather}
\bar{g}^*_{1,0}=\pm\f{\sqrt{\eps}}{\pi}\, ; \quad \bar{g}^*_{2,0}=\f{9}{2\pi}\left(-1\mp 2\sqrt{\eps}\right)\, ; \quad \bar{g}^*_{3,0}=0\, ;
\quad
\bar{g}^*_{4,0}=\f{9}{7\pi}\left(5 \pm 7\sqrt{\eps}\right)\, ;\quad \bar{g}^*_{5,0} = \f{-109\mp 126\sqrt{\eps}}{42\pi}.
\label{eq:FP_LO}
\end{gather}

The signs for all five couplings are taken to be simultaneously either the upper or lower ones. We thus recover the two interacting large-$N$ fixed points of \cite{Benedetti:2019rja}. These fixed points exhibit an interesting new feature: apart from the wheel coupling, all critical couplings start at order $\mathcal{O}(\epsilon^0)$ and not $\mathcal{O}(\epsilon^{1/2})$. This is very different from the quartic model and is due to the fact that the graph $D_1$ contributes at leading order only with wheel vertices. Indeed, all contributions of this graph with other interactions as vertices are of order $1/N$ or higher. Moreover, these fixed points are still perturbative because the beta functions $\beta_i$ for $i \geq 2$ are linear in $\bar{g}_i$ exactly at large $N$. More precisely, they are linear combinations of the couplings, with coefficients that are functions of $\bar{g}_1^2$:
\be
\beta_i = -2\epsilon \bar{g}_i + \tilde{A}_i(\bar{g}_1^2) + \sum_j  \tilde{B}_{ij}(\bar{g}_1^2)  \bar{g}_j \,,
\ee
where $\tilde{A}_i(\bar{g}_1^2)$ and $\tilde{B}_{ij}(\bar{g}_1^2)$ are series in $\bar{g}_1^2$  (see \cite{Benedetti:2019rja} for the complete derivation). The critical coupling $\bar{g}_1^{\star}$ being of order $\sqrt{\epsilon}$ at large $N$, this is indeed a perturbative expansion.

Substituting \eqref{eq:FP_LO} into the order $\tilde{N}^{-1}$ of the beta functions, we find the following corrections to the fixed points:

\begin{gather}
\bar{g}^*_{1,1}= \pm 6 \pi \epsilon^{3/2} , \qquad \bar{g}^*_{2,1}= \frac{9}{4\pi}\left(-1\pm4\sqrt{\epsilon}\right)+\mathcal{O}(\epsilon) , \qquad \bar{g}^*_{3,1}=- \frac{3}{2\pi} +\mathcal{O}(\epsilon), 
\crcr
\bar{g}^*_{4,1}= \frac{9(68-\pi^2)}{98\pi}\mp \frac{9(392+15\pi^2)\sqrt{\epsilon}}{196\pi}+\mathcal{O}(\epsilon) , \crcr
\bar{g}^*_{5,1}= \frac{-18459+566\pi^2}{6860 \pi} \pm \frac{3(2940+193\pi^2)\sqrt{\epsilon}}{980 \pi}+\mathcal{O}(\epsilon) \, .
\end{gather}
where the choice of sign is the same as for the leading order so that we still have two fixed points.

We then compute the critical exponents up to order $\tilde{N}^{-1}$. They are the eigenvalues of the stability matrix $\frac{\partial \beta_i}{\partial \tilde{g}_j}$ evaluated at the fixed points. We find for both fixed points:

\begin{equation}
\left(4\epsilon  \, , \, 4\epsilon-\frac{4\epsilon}{\tilde{N}}\pm\frac{8\epsilon^{3/2}}{\tilde{N}} \, , \,  6\epsilon  \, , \, 14\epsilon-\frac{(16+\pi^2)\epsilon}{2\tilde{N}}\mp\frac{2\pi^2}{\tilde{N}}\epsilon^{3/2} \, , \, 30 \epsilon   \right) \, ,
\end{equation}
where the choice of sign is the same as in \eqref{eq:FP_LO}. The full stability matrix is given in App.~\ref{ap:stability}.

All critical exponents are real positive. Therefore, both fixed points are infrared stable. Moreover, contrary to the large-$N$ case, we now have five different eigenvalues: the stability matrix is diagonalizable at order $\tilde{N}^{-1}$. 

We can finally compute the field critical exponent at the fixed points. We find for both fixed points:

\begin{equation}
\eta(\bar{g}^{\star})=\frac{2\pi^2}{3}\left(\frac{\epsilon}{\pi^2}+\frac{12\epsilon^2}{N}\right)+\mathcal{O}(N^{-2}) \, .
\end{equation}

\section{The long-range sextic multi-scalar model}
\label{sec:MS_LR}

\subsection{Action}

The long-range multi-scalar model with sextic interactions and complex fields in dimension $d$ is defined by the action:

\begin{align}
		S[\phi]  \, &= \, \int d^dx \, \bigg[ \frac{1}{2}  \bar{\phi}_\mba(x)\left(- \partial^2\right)^{\zeta} \phi_{\mba}(x)
		\, + \, \frac{1}{(3!)^2} \, \lambda_{\mba \mbb \mbc; \mbd \mbe \mbf } \phi_{\mba}(x)\phi_{\mbb}(x)\phi_{\mbc}(x) \bar{\phi}_{\mbd}(x)  \bar{\phi}_{\mbe}(x)  \bar{\phi}_{\mbf}(x) \bigg] \, .
	\end{align}
	
This model is called long-range because of the non-trivial power of the Laplacian $0<\zeta<1$. The parameter $\zeta$ must be strictly positive in order to have a well-defined thermodynamic limit\footnote{Models with negative $\zeta$ are still of phenomenological interest. They are called \textit{strong} long-range models by opposition to \textit{weak} long-range models for positive $\zeta$ \cite{Mukamel:2009notes}.}. It is bounded above by $1$ to satisfy reflection positivity.
For this entire section, the dimension is now fixed to be smaller than three (but not necessarily close to three).

The covariance of the free theory is:
\begin{equation}
C(p)=\frac{1}{p^{2\zeta}}=\frac{1}{\Gamma(\zeta)}\int_0^{\infty} da \, a^{\zeta-1}e^{-ap^2} \, ,
\end{equation}
and the canonical dimension of the field is $\Delta_{\phi}=\frac{d-2\zeta}{2}$. This means that for $\zeta<d/3$ the sextic interaction is irrelevant and leads to a mean-field behavior \cite{Aizenman:1988}. On the contrary, for $\zeta>d/3$, the sextic interaction is now relevant and we can expect a non-trivial IR behavior. Finally, at exactly $\zeta=d/3$, we are in the marginal case.

Here, we will use dimensional regularization in the weakly relevant case: $\zeta=\frac{d+\epsilon}{3}$. As for the short-range case we use BPHZ subtraction at zero momentum and introduce an IR regulator by modifying the covariance as:

\begin{equation}
C_{\mu}(p)=\frac{1}{\left(p^2+\mu^2\right)^{\zeta}}=\frac{1}{\Gamma(\zeta)}\int_0^{\infty} da \, a^{\zeta-1}e^{-ap^2-a\mu^2}
\end{equation}
for some mass parameter $\mu>0$.

The key difference with the short-range case is that we now don't have any wave function renormalization. Indeed, the Laplacian is non-local while the divergences are local: it is not renormalized. The bare and renormalized fields thus coincide and there is no anomalous dimension. This is an important feature of long-range models \cite{Fisher:1972zz,suzuki1972wilson,brezin2014crossover,Behan:2017emf,Brydges:2002wq}. In particular, a rigorous proof of the absence of anomalous dimension for the two-point function can be found in \cite{lohmann2017critical}.

\subsection{Beta functions}

The beta function is then the scale derivative of the running coupling at fixed bare coupling:

\begin{equation}
\beta_{\mba \mbb \mbc; \mbd \mbe \mbf}=\mu\partial_{\mu}g_{\mba \mbb \mbc ;\mbd \mbe \mbf}\, .
\end{equation}

The running coupling is defined by:

\begin{equation}
g_{\mba \mbb \mbc; \mbd \mbe \mbf}=\mu^{-2\epsilon}\Gamma^{(6)}_{\mba \mbb \mbc ;\mbd \mbe \mbf}(0,\dots,0) \, ,
\end{equation}
with the following bare expansion:

\begin{align}
&\Gamma^{(6)}_{\mba \mbb \mbc; \mbd \mbe \mbf}(p_1,\dots,p_6)=\lambda_{\mba \mbb \mbc ;\mbd \mbe \mbf}-\frac{1}{6}\left[ 3\left(\lambda_{\mba \mbb \mbg ;\mbh \mbj \mbd}\lambda_{\mbc \mbh \mbj ;\mbe \mbf \mbg}+ 8\text{ terms}\right) +\lambda_{\mba \mbb \mbc ;\mbg \mbh \mbj}\lambda_{\mbg \mbh \mbj ;\mbd \mbe \mbf}\right]\mu^{-2\epsilon}D_{d/3} \crcr
& +\frac{1}{12}\left[3\left(\lambda_{\mba \mbg \mbh ; \mbj \mbk \mbl}\lambda_{\mbb \mbj \mbk ;\mbg \mbh \mbm}\lambda_{\mbl \mbm \mbc ;\mbd \mbe \mbf} + 2\text{ terms}\right)+ 3\left(\lambda_{ \mbj \mbk \mbl ; \mbd \mbg \mbh}\lambda_{ \mbg \mbh \mbm ;\mbe \mbj \mbk}\lambda_{ \mba \mbb \mbc ;\mbf \mbl \mbm} + 2\text{ terms}\right) \right.\crcr
&\left.  \qquad + 3\left(\lambda_{\mba \mbg \mbh ;\mbj \mbk \mbl}\lambda_{ \mbj \mbk \mbm ;\mbg \mbh \mbd}\lambda_{\mbl \mbb \mbc ;\mbe \mbf \mbm} + 8\text{ terms}\right)+2\left(\lambda_{\mba \mbg \mbh ;\mbj \mbk \mbl }\lambda_{ \mbj \mbk \mbl ;\mbh \mbm \mbd}\lambda_{\mbm \mbb \mbc ;\mbe \mbf \mbg} + 8\text{ terms}\right)\right]\mu^{-4\epsilon}S_{d/3} \crcr
& +\frac{1}{36}\left[ 9\left(\lambda_{\mba \mbg \mbh ;\mbj \mbe \mbf}\lambda_{\mbj \mbk \mbl ;\mbm \mbg \mbh}\lambda_{\mbm \mbb \mbc ;\mbk \mbl \mbd}+ 8\text{ terms}\right) +\lambda_{\mba \mbb \mbc ;\mbg \mbh \mbj}\lambda_{\mbg \mbh \mbj ;\mbk \mbl \mbm}\lambda_{\mbk \mbl \mbm ;\mbd \mbe \mbf}\right]\mu^{-4\epsilon}D_{d/3}^2 \crcr
& +\frac{1}{4}\left[\left(\lambda_{\mba \mbg \mbh ; \mbd \mbj \mbk}\lambda_{\mbb \mbc \mbl ;\mbg  \mbh \mbm}\lambda_{\mbj  \mbk \mbm ; \mbl \mbe \mbf} + 8 \text{ terms}\right)+ 2\left(\lambda_{\mba \mbg  \mbh ;\mbd  \mbj \mbk }\lambda_{\mbb \mbc \mbj; \mbg \mbl \mbm}\lambda_{\mbk  \mbl \mbm ;\mbh \mbe \mbf} + 8 \text{ terms}\right) \right. \crcr
& \left. \qquad +\left(\lambda_{\mba \mbg \mbh ;\mbd  \mbj \mbk}\lambda_{\mbb \mbl \mbm ;\mbe \mbg \mbh}\lambda_{\mbc  \mbj \mbk ;\mbf \mbl \mbm} + 5 \text{ terms}\right)+4\left(\lambda_{\mba \mbg \mbh ; \mbd  \mbj \mbk}\lambda_{\mbb \mbj \mbl ;\mbe \mbg \mbm}\lambda_{\mbc  \mbk \mbm ;\mbf \mbh \mbl} + 5 \text{ terms}\right)\right]\mu^{-4\epsilon}I_{d/3} \crcr
& +\frac{1}{12}\left[\left(\lambda_{\mba \mbg \mbh ;\mbj \mbe \mbf }\lambda_{\mbb \mbc \mbj ;\mbk \mbl \mbm }\lambda_{\mbk \mbl \mbm ; \mbg \mbh   \mbd} +  \text{ 8 terms}\right) + 6\left(\lambda_{\mba \mbg \mbh ;\mbj \mbe \mbf }\lambda_{\mbb \mbc \mbk ;\mbg \mbl \mbm }\lambda_{\mbl \mbm\mbj ;\mbh \mbk  \mbd} +  \text{ 8 terms}\right)\right. \crcr
& \qquad + \left(\lambda_{\mbj \mbb \mbc ;\mbd \mbg \mbh }\lambda_{\mbk \mbl \mbm ;\mbe \mbf \mbj  }\lambda_{\mba \mbg \mbh ;\mbk \mbl \mbm} +  \text{ 8 terms}\right) + 6\left(\lambda_{\mbj \mbb \mbc ;\mbd  \mbg \mbh }\lambda_{\mbg \mbl \mbm ;\mbk \mbe \mbf }\lambda_{\mba \mbh \mbk ;\mbl  \mbm \mbj} +  \text{ 8 terms}\right)\crcr
& \qquad +3\left(\lambda_{\mba \mbk \mbl ;\mbm \mbe \mbd}\lambda_{\mbb \mbj \mbm  ;\mbf \mbg  \mbh }\lambda_{\mbc \mbg \mbh ; \mbj  \mbk \mbl} +  \text{ 17 terms}\right)+6\left(\lambda_{\mba \mbl \mbm ;\mbh \mbd \mbe }\lambda_{\mbb  \mbj \mbk ;\mbg\mbm  \mbf }\lambda_{\mbc \mbg \mbh ;\mbj  \mbk \mbl} +  \text{ 17 terms}\right)\crcr
& \qquad +3\left(\lambda_{ \mbm \mba \mbb ;\mbd \mbk \mbl}\lambda_{\mbc   \mbg \mbh ;\mbe \mbj  \mbm}\lambda_{\mbj \mbk \mbl  ;\mbg \mbh   \mbf} +  \text{ 17 terms}\right)+6\left(\lambda_{ \mba \mbb  \mbh ;\mbd \mbl\mbm}\lambda_{\mbc   \mbg \mbm ;\mbe \mbj   \mbk}\lambda_{\mbj \mbk \mbl ;\mbg \mbh   \mbf} +  \text{ 17 terms}\right) \crcr
& \qquad \left. +3\left(\lambda_{\mba \mbb \mbc ;\mbg \mbh \mbj}\lambda_{\mbj \mbl \mbm ;\mbk \mbd \mbe}\lambda_{\mbg \mbh \mbk ;\mbl \mbm \mbf} + \text{ 2 terms}\right)+ 3\left(\lambda_{\mbg \mbh \mbj ;\mbd \mbe \mbf}\lambda_{ \mba \mbb \mbk ;\mbj \mbl \mbm}\lambda_{ \mbl \mbm \mbc ;\mbg \mbh \mbk }+ \text{ 2 terms}\right)\right]\mu^{-4\epsilon}J_{d/3} \, .
\end{align}

We rescale the couplings by $\tilde{g}_{\mba \mbb \mbc ;\mbd \mbe \mbf}=(4\pi)^{-d}\Gamma(d/3)^{-3}g_{\mba \mbb \mbc ; \mbd \mbe \mbf}$ and we define:

\begin{align}
\alpha_{D_{d/3}}&=\epsilon(4\pi)^d\Gamma(d/3)^{3}\frac{D_{d/3}}{3}=\frac{\Gamma(d/6)^3}{3\Gamma(d/2)} \; , \; \alpha_{S_{d/3}}=-\epsilon(4\pi)^{2d} \Gamma(d/3)^{6}\frac{S_{d/3}}{3}= -\frac{\Gamma(d/6)^4\Gamma(-d/6)\Gamma(d/3)^2}{6\Gamma(d/2)\Gamma(2d/3)}\crcr
\alpha_{I_{d/3}}&=-\epsilon(4\pi)^{2d}\Gamma(d/3)^{6}I_{d/3}= -\frac{\Gamma(d/6)^9}{2\Gamma(d/3)^3\Gamma(d/2)}\; ,  \crcr
\alpha_{J_{d/3}}&=\epsilon(4\pi)^{2d}\Gamma(d/3)^{6}\frac{(D_{d/3}^2-2J_{d/3})}{6}= \frac{\Gamma(d/6)^6}{6\Gamma(d/2)^2\Gamma(d/3)^6}\Bigg[\psi(d/6)-\psi(1)+\psi(d/3)-\psi(d/2)\Bigg] \, .
\end{align}

We finally obtain the following beta functions:
\begin{align}
\beta_{\mba \mbb \mbc ;\mbd \mbe \mbf}&=-2\epsilon \tilde{g}_{\mba \mbb \mbc ;\mbd \mbe \mbf}+\alpha_{D_{d/3}}\left[ 3\left(\tilde{g}_{\mba \mbb \mbg ;\mbh \mbj \mbd}\tilde{g}_{\mbc \mbh \mbj ;\mbe \mbf \mbg}+ 8\text{ terms}\right) +\tilde{g}_{\mba \mbb \mbc ;\mbg \mbh \mbj}\tilde{g}_{\mbg \mbh \mbj ;\mbd \mbe \mbf}\right] \crcr
& +\alpha_{S_{d/3}}\left[3\left(\tilde{g}_{\mba \mbg \mbh ; \mbj \mbk \mbl}\tilde{g}_{\mbb \mbj \mbk ;\mbg \mbh \mbm}\tilde{g}_{\mbl \mbm \mbc ;\mbd \mbe \mbf} + 2\text{ terms}\right)+ 3\left(\tilde{g}_{ \mbj \mbk \mbl ; \mbd \mbg \mbh}\tilde{g}_{ \mbg \mbh \mbm ;\mbe \mbj \mbk}\tilde{g}_{ \mba \mbb \mbc ;\mbf \mbl \mbm} + 2\text{ terms}\right) \right.\crcr
&\left.  \qquad + 3\left(\tilde{g}_{\mba \mbg \mbh ;\mbj \mbk \mbl}\tilde{g}_{ \mbj \mbk \mbm ;\mbg \mbh \mbd}\tilde{g}_{\mbl \mbb \mbc ;\mbe \mbf \mbm} + 8\text{ terms}\right)+2\left(\tilde{g}_{\mba \mbg \mbh ;\mbj \mbk \mbl }\tilde{g}_{ \mbj \mbk \mbl ;\mbh \mbm \mbd}\tilde{g}_{\mbm \mbb \mbc ;\mbe \mbf \mbg} + 8\text{ terms}\right)\right] \crcr
& +\alpha_{I_{d/3}}\left[\left(\tilde{g}_{\mba \mbg \mbh ; \mbd \mbj \mbk}\tilde{g}_{\mbb \mbc \mbl ;\mbg  \mbh \mbm}\tilde{g}_{\mbj  \mbk \mbm ; \mbl \mbe \mbf} + 8 \text{ terms}\right)+ 2\left(\tilde{g}_{\mba \mbg  \mbh ;\mbd  \mbj \mbk }\tilde{g}_{\mbb \mbc \mbj; \mbg \mbl \mbm}\tilde{g}_{\mbk  \mbl \mbm ;\mbh \mbe \mbf} + 8 \text{ terms}\right) \right. \crcr
& \left. \qquad +\left(\tilde{g}_{\mba \mbg \mbh ;\mbd  \mbj \mbk}\tilde{g}_{\mbb \mbl \mbm ;\mbe \mbg \mbh}\tilde{g}_{\mbc  \mbj \mbk ;\mbf \mbl \mbm} + 5 \text{ terms}\right)+4\left(\tilde{g}_{\mba \mbg \mbh ; \mbd  \mbj \mbk}\tilde{g}_{\mbb \mbj \mbl ;\mbe \mbg \mbm}\tilde{g}_{\mbc  \mbk \mbm ;\mbf \mbh \mbl} + 5 \text{ terms}\right)\right]\crcr
& +\alpha_{J_{d/3}}\left[\left(\tilde{g}_{\mba \mbg \mbh ;\mbj \mbe \mbf }\tilde{g}_{\mbb \mbc \mbj ;\mbk \mbl \mbm }\tilde{g}_{\mbk \mbl \mbm ; \mbg \mbh   \mbd} +  \text{ 8 terms}\right) + 6\left(\tilde{g}_{\mba \mbg \mbh ;\mbj \mbe \mbf }\tilde{g}_{\mbb \mbc \mbk ;\mbg \mbl \mbm }\tilde{g}_{\mbl \mbm\mbj ;\mbh \mbk  \mbd} +  \text{ 8 terms}\right)\right. \crcr
& \qquad \left(\tilde{g}_{\mbj \mbb \mbc ;\mbd \mbg \mbh }\tilde{g}_{\mbk \mbl \mbm ;\mbe \mbf \mbj  }\tilde{g}_{\mba \mbg \mbh ;\mbk \mbl \mbm} +  \text{ 8 terms}\right) + 6\left(\tilde{g}_{\mbj \mbb \mbc ;\mbd  \mbg \mbh }\tilde{g}_{\mbg \mbl \mbm ;\mbk \mbe \mbf }\tilde{g}_{\mba \mbh \mbk ;\mbl  \mbm \mbj} +  \text{ 8 terms}\right)\crcr
& \qquad +3\left(\tilde{g}_{\mba \mbk \mbl ;\mbm \mbe \mbd}\tilde{g}_{\mbb \mbj \mbm  ;\mbf \mbg  \mbh }\tilde{g}_{\mbc \mbg \mbh ; \mbj  \mbk \mbl} +  \text{ 17 terms}\right)+6\left(\tilde{g}_{\mba \mbl \mbm ;\mbh \mbd \mbe }\tilde{g}_{\mbb  \mbj \mbk ;\mbg\mbm  \mbf }\tilde{g}_{\mbc \mbg \mbh ;\mbj  \mbk \mbl} +  \text{ 17 terms}\right)\crcr
& \qquad +3\left(\tilde{g}_{ \mbm \mba \mbb ;\mbd \mbk \mbl}\tilde{g}_{\mbc   \mbg \mbh ;\mbe \mbj  \mbm}\tilde{g}_{\mbj \mbk \mbl  ;\mbg \mbh   \mbf} +  \text{ 17 terms}\right)+6\left(\tilde{g}_{ \mba \mbb  \mbh ;\mbd \mbl\mbm}\tilde{g}_{\mbc   \mbg \mbm ;\mbe \mbj   \mbk}\tilde{g}_{\mbj \mbk \mbl ;\mbg \mbh   \mbf} +  \text{ 17 terms}\right) \crcr
& \qquad \left. +3\left(\tilde{g}_{\mba \mbb \mbc ;\mbg \mbh \mbj}\tilde{g}_{\mbj \mbl \mbm ;\mbk \mbd \mbe}\tilde{g}_{\mbg \mbh \mbk ;\mbl \mbm \mbf} + \text{ 2 terms}\right)+ 3\left(\tilde{g}_{\mbg \mbh \mbj ;\mbd \mbe \mbf}\tilde{g}_{ \mba \mbb \mbk ;\mbj \mbl \mbm}\tilde{g}_{ \mbl \mbm \mbc ;\mbg \mbh \mbk }+ \text{ 2 terms}\right)\right] \, .
\label{eq:beta_LR}
\end{align}

\subsection{Application: $U(N)^3$ symmetry}
\label{sec:tensor_LR}

We specify again the symmetry to $U(N)^3$ as in section \ref{sec:tensor_SR}. Now, for the long-range case, setting $\zeta=\frac{d+\epsilon}{3}$, we obtain the following beta functions up to order $N^{-1}$:
\begin{align}
\beta_1&=-2\epsilon \bar{g}_1+\frac{48}{N}\alpha_{I_{d/3}}\bar{g}_1^3 +\mathcal{O}(N^{-2})\, ,\crcr
\beta_2&=-2\epsilon\bar{g}_2+3\bar{g}_1^2\left(9\alpha_{D_{d/3}}+2\alpha_{S_{d/3}}(9\bar{g}_1+\bar{g}_2)\right) +\frac{2}{N}\Bigg[\frac{\alpha_{D_{d/3}}}{3}\left(81\bar{g}_1^2+36\bar{g}_1\bar{g}_2+\bar{
g}_2^2+6\bar{g}_3(9\bar{g}_1+\bar{g}_2)\right)\crcr
& \qquad +2\bar{g}_1^2\left(9(4\alpha_{J_{d/3}}+\alpha_{S_{d/3}})(3\bar{g}_1+\bar{g}_3)+2(4\alpha_{I_{d/3}}+3\alpha_{S_{d/3}})\bar{g}_2\right)\Bigg]+\mathcal{O}(N^{-2})\, , \crcr
\beta_3&=-2\epsilon\bar{g}_3+9\alpha_{S_{d/3}}\bar{g}_1^2\bar{g}_3 \crcr
& \quad +\frac{1}{N}\Bigg[3\alpha_{D_{d/3}}\bar{g}_3^2+\frac{2\alpha_{D_{d/3}}}{3}\bar{g}_2\left(18\bar{g}_1+\bar{g}_2\right)+108\alpha_{J_{d/3}}\bar{g}_1^3+12\alpha_{S_{d/3}}\bar{g}_1^2\bar{g}_2\Bigg]+\mathcal{O}(N^{-2})\, ,\crcr
\beta_4&=-2\epsilon\bar{g}_4+3\alpha_{S_{d/3}}\bar{g}_1^2\left(27\bar{g}_1+10\bar{g}_2+12\bar{g}_3+7\bar{g}_4\right)\crcr
& \quad +\frac{1}{27N}\Bigg[9\alpha_{D_{d/3}}\left(36\bar{g}_1\left(\bar{g}_2+3\bar{g}_3+2\bar{g}_4\right)+2\bar{g}_2\left(5\bar{g}_2+12\bar{g}_3+4\bar{g}_4\right)+3\bar{g}_3\left(9\bar{g}_3+4\bar{g}_4\right)\right) \crcr
& \qquad +\frac{2}{3}\alpha_{I_{d/3}}\big(162\bar{g_1}^2\left(54\bar{g}_1+5\bar{g}_2+12\bar{g}_3+4\bar{g}_4\right)+2\bar{g}_2^2\left(2\bar{g}_2+\bar{g}_4\right)+3\bar{g}_3\left(2\bar{g}_2^2+12\bar{g}_2\bar{g}_3+9\bar{g}_3^2+6\bar{g}_3\bar{g}_4\right)  \big)\crcr
& \qquad +324\alpha_{S_{d/3}}\bar{g}_1^2\left(4\bar{g}_2+6\bar{g}_3+\bar{g}_4\right)+81\alpha_{J_{d/3}}\bar{g}_1^2\left(36\bar{g}_1+33\bar{g}_2+18\bar{g}_3+8\bar{g}_4\right) \Bigg] +\mathcal{O}(N^{-2})\, ,\crcr
\beta_5&=-2\epsilon\bar{g}_5+3\bar{g}_1^2\left(\alpha_{D_{d/3}}+\alpha_{S_{d/3}}(3\bar{g}_2+8\bar{g}_4+15\bar{g}_5)\right) \crcr
& \quad +\frac{1}{N}\Bigg[\frac{2}{3}\alpha_{D_{d/3}}\left(\bar{g}_2(\bar{g}_2+6\bar{g}_3+4\bar{g}_4)+2\bar{g}_4(3\bar{g}_3+\bar{g}_4)\right)  +6\alpha_{S_{d/3}}\bar{g}_1^2\left(3\bar{g}_3+4\bar{g}_4\right) \crcr
& \qquad +\frac{4}{243}\alpha_{I_{d/3}}\big(243\bar{g_1}^2\left(\bar{g}_2+3\bar{g}_3+\bar{g}_4\right)+10\bar{g}_2^3+\bar{g}_4\left(3\bar{g}_2\left(7\bar{g}_2+24\bar{g}_3+4\bar{g}_4\right)+9\bar{g}_3\left(3\bar{g}_3+2\bar{g}_4\right)+2\bar{g}_4\right) \crcr
&  \qquad \qquad +9\bar{g}_3\left(7\bar{g}_2^2+6\bar{g}_2\bar{g}_3\right)  \big) +3\alpha_{J_{d/3}}\bar{g}_1^2\left(3\bar{g}_2+6\bar{g}_3+8\bar{g}_4\right) \Bigg]+\mathcal{O}(N^{-2})\, .
\end{align}

In the long-range case, at $\epsilon=0$, the wheel coupling $\bar{g}_1$ is exactly marginal. However, at order $N^{-1}$ the wheel beta function is non zero and the line of fixed points found in \cite{Benedetti:2019rja} collapses to the trivial fixed point. 
Turning on $\epsilon$ does not solve the problem as it contributes a term $-2\epsilon\bar{g}_1$ which already gives $\bar{g}_1^{\star}=0$ at leading order. As for the short-range case, we should also consider how small $\epsilon$ is compared to $N$. At next-to-leading order the wheel beta function has the following form $-2\epsilon \bar{g}_1+\bar{g}_1^3/N$. Its fixed points are the trivial one and $\bar{g}_1^{\star}=\sqrt{N\epsilon}$. The latter goes to infinity for $N \rightarrow \infty$ at fixed $\epsilon$. This is solved by imposing $N\epsilon \ll 1$. We thus set:
\begin{equation}
\epsilon=\frac{\tilde{\epsilon}}{N} \, ,
\end{equation}
and as before we expand first in $1/N$ and then in $\tilde{\epsilon}$. This is again similar to what happens in \cite{Benedetti:2020sye}.

We parametrize again the critical couplings as $\bar{g}_i=\bar{g}_{i,0}+\frac{\bar{g}_{i,1}}{N} +\mathcal{O}(N^{-2})$ for $i=1,\dots , 5$. Solving for the zeros of the beta functions at leading order we find the following solutions\footnote{There is also a solution with zero wheel coupling leading to a $4$-dimensional manifold of fixed points. We do not study this solution further as we are only interested in solutions with non-zero wheel coupling in order to have a melonic fixed point.}:

\begin{gather}
\bar{g}_{2,0}^*=-9\bar{g}_{1,0}+\f{9\Gamma(2d/3)}{\Gamma(-d/6)\Gamma(d/6)\Gamma(d/3)^2},\qquad \bar{g}_{3,0}^*=0,\crcr
\bar{g}_{4,0}^*=9\bar{g}_{1,0} - \frac{90}{7}\f{\Gamma(2d/3)}{\Gamma(-d/6)\Gamma(d/6)\Gamma(d/3)^2},\crcr
\bar{g}_{5,0}^*=-3\bar{g}_{1,0} + \f{109\Gamma(2d/3)}{21\Gamma(-d/6)\Gamma(d/6)\Gamma(d/3)^2}\, .
\end{gather}

We thus again recover the line of fixed points found in \cite{Benedetti:2019rja}.
Solving the wheel beta functions at order $N^{-1}$, we find:
\begin{equation}
\bar{g}_{1,0}=\pm \frac{ \sqrt{\tilde{\epsilon}}}{2\sqrt{6\alpha_{I_{d/3}}}} \;.
\end{equation}

$I_{d/3}$ being negative, $\bar{g}_{1,0}$ is thus purely imaginary. This implies that the other four couplings are also complex at leading order. 
However, substituting these results into the order $N^{-1}$ of the beta functions, we find non perturbative corrections to the fixed points which blow up when sending $\tilde{\epsilon} \rightarrow 0$. This cannot be fixed by rescaling $\tilde{\epsilon}$ or $N$. 
This is due to the form of the beta functions. Indeed, as we saw in the short-range model, all couplings except the wheel start at order $0$ in $\epsilon$. When solving at next-to-leading order, this will lead to non-perturbative results. To cure this, we could rescale $N$ as we did for the short-range model, $\tilde{N}=\epsilon N$. However, doing so instead of rescaling $\epsilon$, the only fixed point is the trivial one.\footnote{Solving at leading order first we find all critical couplings to be equal to zero $\bar{g}_{i,0}^{\star}=0$. Re-injecting this solution into the next-to-leading order of the beta functions, they reduce for all couplings to $\beta_{i,1}=-2\epsilon \bar{g}_{i,1}$. This indeed leads to only a trivial fixed point.} 
We therefore conclude that there is no precursor at next-to-leading order of the large-$N$ fixed point.

%
%
%

\section{Conclusion}

In this paper we studied bosonic tensor models with sextic interactions at next-to-leading order. We considered only the model of rank $3$ of \cite{Benedetti:2019rja} as only trivial fixed points were found in rank $5$ at large $N$. We chose as free propagator either the standard short-range propagator or the critical long-range propagator. In both cases, we studied the renormalization group and computed fixed points of the beta functions at next-to-leading order. However, the results are radically different in the two cases. 
In the short-range case, the theory admits a non-trivial real stable IR fixed point with non-zero wheel coupling, thus leading to melonic dominance.  
In the long-range case, the corrections to the large-$N$ fixed points are not perturbative in $\tilde{\epsilon}$ and even blow up when $\tilde{\epsilon}$ goes to zero. This indicates that the large-$N$ fixed point found in \cite{Benedetti:2019rja} has no precursor at next-to-leading order.

As for the computation of $1/N$ corrections in quartic models \cite{Benedetti:2020sye}, a subtle part of our analysis is the identification of a proper hierarchy between our two small parameters $1/N$ and $\epsilon$. Indeed, in the short-range case we need $\epsilon N \gg 1$ while in the long-range case we need $\epsilon N \ll 1$. These conditions are found by demanding that the large-$N$ fixed points remain dominant in the beta functions. However, for the long-range case, contrary to quartic models, this condition is not enough to ensure a perturbative solution of the beta functions at next-to-leading order. This is due to the presence of a term of order $\mathcal{O}(\epsilon^0)$ in the large-$N$ fixed point.

Nevertheless, this is an interesting new feature of our fixed point that also appears in the short-range case. Indeed, the wheel coupling at large $N$ is of order $\sqrt{\epsilon}$ while the other couplings start at order $\epsilon^0$. This is very different from usual Wilson-Fisher like fixed points and from the quartic model fixed points \cite{Benedetti:2020sye}. It is due to the fact that the graph $D_1$ contributes to leading order in $N$ only with wheel vertices whereas in quartic models the one-loop Feynman graph contributes to leading order with all three quartic interactions on the vertices. This model thus leads to a new kind of melonic fixed point. 

One other interesting feature we found is the diagonalizability of the stability matrix at next-to-leading order in the short-range case. At large $N$, the stability matrix was non diagonalizable due to an eigenvalue of multiplicity two whereas at next-to-leading order, we have five different eigenvalues and the stability matrix is diagonalizable. This suggests that the logarithmic CFT of the large-$N$ limit reduces to an ordinary CFT at next-to-leading order. However, this statement requires further investigation that we leave for future work. 

\section*{Acknowledgements}

I would like to thank Dario Benedetti and Razvan Gurau for helpful discussions and thorough reviews of the draft. I would also like to thank Nicolas Delporte for useful discussions. 
The work of SH is supported by the European Research Council (ERC) under the European Union's Horizon 2020 research and innovation program (grant agreement No818066) and by Deutsche Forschungsgemeinschaft (DFG, German Research Foundation) under Germany's Excellence Strategy EXC-2181/1 - 390900948 (the Heidelberg STRUCTURES Cluster of Excellence)

\appendix

\section{The renormalized series}
\label{ap:renor}

To compute the beta functions in practice, we can derive the bare expansion \eqref{eq:bare_series} with respect to $\mu$ and then substitute the bare constants in terms of the renormalized ones using the renormalized series. 
The renormalized series can be obtained by explicit computation or immediately using the Bogoliubov Parasuk recursion as in appendix A of \cite{Benedetti:2020rrq}. 

For our multi-scalar model, both in short and long range we have at order two in the renormalized coupling:

\begin{equation}
\mu^{-2\epsilon}\lambda_{\mba \mbb \mbc ;\mbd \mbe \mbf}=g_{\mba \mbb \mbc ;\mbd \mbe \mbf}+\frac{D_{\zeta}}{6}\left[ 3\left(\lambda_{\mba \mbb \mbg ;\mbh \mbj \mbd}\lambda_{\mbc \mbh \mbj ;\mbe \mbf \mbg}+ 8\text{ terms}\right) +\lambda_{\mba \mbb \mbc ;\mbg \mbh \mbj}\lambda_{\mbg \mbh \mbj ;\mbd \mbe \mbf}\right]\, .
\end{equation}

This then allows us to obtain the beta functions of \eqref{eq:beta_SR} and \eqref{eq:beta_LR}.

\section{The melon integral}
\label{ap:melon}

In this section we compute the melon integral contributing to the wave function renormalization in the short-range case:
\begin{equation}
M_{1}(p)=\int_{q_1,q_2,q_3,q_4}G_0(q_1)G_0(q_2)G_0(q_3)G_0(q_4)G_0(p+q_1+q_2+q_3+q_4) \,,
\end{equation}
with $G_0(p)=\frac{1}{p^{2}}$.

We will use the following formula:
\begin{equation}
\int\f{\dd[d] k}{(2\pi)^d} \f{1}{k^{2\a}(k+p)^{2\b}}  = \f{1}{(4\pi)^{d/2}}\f{\G(\tfrac{d}{2} - \a)\G(\tfrac{d}{2}  - \b)\G(\a +\b -\tfrac{d}{2} )}{\G(\a)\G(\b) \G(d - \a - \b)}\f{1}{|p|^{2(\a+\b - \tfrac{d}{2} )}} \,.
\label{eq:intG}
\end{equation}
We obtain:
\begin{equation}
M_{1}(p)=\frac{p^{4d-10}}{(4\pi)^{2d}}\frac{\Gamma(\tfrac{d}{2} -1)^5\Gamma(5-2d)}{\Gamma(\tfrac{5d}{2} -5)} \,.
\end{equation}
For $d=3-\epsilon$, this simplifies to:
\begin{equation}  \label{eq:M-1}
M_1(p)=\frac{p^{2-4\epsilon}}{(4\pi)^{6-2\epsilon}}\frac{\Gamma(2\epsilon-1)\Gamma(\tfrac{1-\epsilon}{2})^5}{\Gamma(\tfrac{5}{2}(1-\epsilon))}\,.
\end{equation}
At first order in $\epsilon$, we finally have:
\begin{equation}
M_1(p)=- \frac{p^{2-4\epsilon}}{(4\pi)^{6}}\frac{2\pi^2}{3\epsilon} + \mathcal{O}(\epsilon^0)\,.
\end{equation}

\section{Beta functions details}
\label{ap:betafun4}

\subsection{2-loop amplitude}
\label{app:D}

We want to compute the two-loop amputated Feynman integral $D_{\zeta}$ represented in Fig.~\ref{fig:graphs}. Because this amplitude appears squared in the coefficient $\alpha_{J_{\zeta}}$ of the beta functions, we need to compute it up to order $\mathcal{O}(1)$. However, at this order, the amplitude is not independent of the choice of IR regularization. Therefore, we have to be careful and compute both $D_{\zeta}$ and $J_{\zeta}$ using the same IR regulator in order to obtain the correct cancellations. As the Gell-Mann and Low subtraction used to compute $D_{\zeta}$ in \cite{Benedetti:2019rja} did not lead to analytical results for $J_{\zeta}$, we will resort here to subtraction at zero momentum for both amplitudes. 

Applying \eqref{eq:amp_final}, the amplitude $D_{\zeta}$ can be written as:
\begin{equation}
D_{\zeta}=\frac{1}{\Gamma(\zeta)^3(4\pi)^d}\int da_1da_2da_3 \frac{(a_1a_2a_3)^{\zeta-1}}{\left(a_1a_2+a_1a_3+a_2a_3\right)^{\tfrac{d}{2}}}e^{-(a_1+a_2+a_3)}\,.
\end{equation}

To obtain this expression, we first associated a Schwinger parameter $a_i$ to each edge in $D_{\zeta}$ and then applied \eqref{eq:amp_final}. To determine the denominator in the integrand, we noticed that this graph has three spanning trees, each consisting of one edge. Equivalently, this expression can be found by writing each propagator with a Schwinger parametrization as in \eqref{eq:free_cov} and integrating the resulting Gaussian integrals on the internal momenta. 

In the following we will repeatedly use the Mellin-Barnes representation for $Re(u)>0$:
\begin{align}
&\frac{1}{\left(A_1+\dots A_{q+1}\right)^u}= \crcr
& \qquad \int_{0^- -i \infty}^{0^- +i\infty} [dz]\frac{\Gamma(-z_1)\dots \Gamma(-z_q)\Gamma(z_1+\dots z_q +u)}{\Gamma(u)}A_1^{z_1}\dots A_q^{z_q}A_{q+1}^{-z_1\dots -z_q-u} \, ,
\label{eq:mellin}
\end{align}
where we denote $[dz]=\frac{dz}{2\pi i}$. The only restriction on the integration contour is $Re(z_i)<0$ (see Appendix B of \cite{Benedetti:2020rrq} for more details).

Using this formula we can rewrite the denominator of $D_{\zeta}$ as:

\begin{equation}
\frac{1}{\left(a_1a_2+a_1a_3+a_2a_3\right)^{\tfrac{d}{2}}}=\int_{0^-} [dz] \frac{\Gamma(-z)\Gamma(z+\tfrac{d}{2})}{\Gamma(\tfrac{d}{2})}\frac{(a_2a_3)^z}{(a_1(a_2+a_3))^{z+\tfrac{d}{2}}} \, .
\end{equation}

We can then integrate the Schwinger parameters using the following formula:

\begin{equation}
\int da_1 da_2 \frac{(a_1a_2)^{u-1}}{(a_1+a_2)^{\gamma}}=\frac{\Gamma(u)^2\Gamma(2u-\gamma)}{\Gamma(2u)}\,,
\label{eq:int_class}
\end{equation}
which is valid for $Re(u)>0$ and $Re(2u)>Re(\gamma)$.

\paragraph{Long-range: $\zeta=\frac{d+\epsilon}{3}$, $d<3$.}

We then obtain:
\begin{equation}
D_{d/3}=\f{1}{(4\pi)^{d}\Gamma(\zeta)^3\Gamma(\tfrac{d}{2})}\int_{-\tfrac{d}{6}^-} [dz] \Gamma(-z)\Gamma(z+\tfrac{d}{2})\frac{\Gamma(\tfrac{d+\epsilon}{3}+z)^2\Gamma(\tfrac{d}{6}+\tfrac{2\epsilon}{3}+z)}{\Gamma(\tfrac{2(d+\epsilon)}{3}+2z)}\Gamma(-\tfrac{d}{6}+\tfrac{\epsilon}{3}-z)\, .
\end{equation}

There is only one pole giving a singularity in $\epsilon$ located at $z=-\frac{d}{6}+\frac{\epsilon}{3}$. 
We thus obtain:
\begin{align}
D_{d/3}=&\f{1}{(4\pi)^{d}\Gamma(\zeta)^3\Gamma(\tfrac{d}{2})}\Big[\frac{\Gamma(\tfrac{d}{6}-\tfrac{\epsilon}{3})\Gamma(\tfrac{d+\epsilon}{3})\Gamma(\tfrac{d}{6}+\tfrac{2\epsilon}{3})^2\Gamma(\epsilon)}{\Gamma(\tfrac{d}{3}+\tfrac{4\epsilon}{3})} \crcr
& \qquad \qquad +\int_{0^-} [dz] \Gamma(-z)\Gamma(z+\tfrac{d}{2})\frac{\Gamma(\tfrac{d}{3}+z)^2\Gamma(\tfrac{d}{6}+z)}{\Gamma(\tfrac{2d}{3}+2z)}\Gamma(-\tfrac{d}{6}-z)\Big]+ \mathcal{O}(\epsilon)\, .
\end{align}

The remaining integral, that we denote $K$, has two types of poles, situated at $z=n_1$ for $n_1\geq 0$ and $z=-\frac{d}{6}+n_2$ for $n_2\geq 1$. We have:
\begin{align}
K=&\sum_{n=0}^{\infty} \frac{(-1)^n}{n!}\frac{\Gamma(n+\tfrac{d}{2})\Gamma(n+\tfrac{d}{3})^2\Gamma(n+\tfrac{d}{6})\Gamma(-\tfrac{d}{6}-n)}{\Gamma(\tfrac{2d}{3}+2n)}+ \sum_{n=1}^{\infty} \frac{(-1)^n}{n!}\frac{\Gamma(\tfrac{d}{6}-n)\Gamma(n+\tfrac{d}{3})\Gamma(n+\tfrac{d}{6})^2}{\Gamma(2n+\tfrac{d}{3})} \, .
\end{align}
Both sums are convergent and can be expressed in terms of hypergeometric functions.\footnote{In particular we have:
\begin{equation*}
K=\frac{\Gamma(-\tfrac{d}{6})\Gamma(\tfrac{d}{3})^3\Gamma(\tfrac{d}{2})}{\Gamma(\tfrac{d}{6})^2\Gamma(\tfrac{2d}{3})}{}_3F_2\left(\tfrac{d}{6},\tfrac{d}{3},\tfrac{d}{2};1+\tfrac{d}{6},\tfrac{1}{2}+\tfrac{d}{3};\tfrac{1}{4}\right) +\frac{d^2}{2(d+3)(6-d)}{}_4F_3\left(1,1,1+\tfrac{d}{6},1+\tfrac{d}{3};2,2-\tfrac{d}{6},\tfrac{3}{2}+\tfrac{d}{6};\tfrac{1}{4}\right) \, ,
\end{equation*}
where ${}_p F_q(a_1 \dots a_p ; b_1 \dots b_q ; z)$ are the generalized hypergeometric functions. This was obtained using Mathematica.} We finally obtain:

\begin{align}
D_{d/3}=\frac{\Gamma(\tfrac{d}{6})^3}{(4\pi)^d \Gamma(\tfrac{d}{3})^3\Gamma(\tfrac{d}{2})}&\left[ \frac{1}{\epsilon}+\psi(1)+\psi(\tfrac{d}{6})-2\psi(\tfrac{d}{3}) + K \right]+\mathcal{O}(\epsilon)\, . 
\end{align}

\paragraph{Short-range: $\zeta=1$, $d=3-\epsilon$.}

Following the same steps, we obtain in the short-range case:
\begin{equation}
D_1=\frac{\pi}{(4\pi)^3}\left[\frac{2}{\epsilon}+\psi(\tfrac{1}{2})+\psi(\tfrac{3}{2})+4\ln(\tfrac{2}{3})\right] +\mathcal{O}(\epsilon) \, .
\end{equation}

\subsection{4-loop amplitude}

\subsubsection{$S_{\zeta}$ integral}


The $S_{\zeta}$ integral with Schwinger parameters can be written as:

\begin{equation}
S_{\zeta}=\frac{1}{\Gamma(\zeta)^6(4\pi)^{2d}}\int da_1da_2db_1db_2db_3db_4 \frac{(a_1a_2b_1b_2b_3b_4)^{\zeta-1}e^{-(a_1+a_2+b_1+b_2+b_3+b_4)}}{\left((a_1+a_2)(b_1b_2(b_3+b_4)+b_3b_4(b_1+b_2))+b_1b_2b_3b_4\right)^{\tfrac{d}{2}}}\, .
\end{equation}

Doing the change of variables $a_1=\alpha \beta$ and $a_2=\alpha(1-\beta)$, we can integrate $\beta$ to obtain:

\begin{equation}
S_{\zeta}=\frac{1}{\Gamma(\zeta)^4\Gamma(2\zeta)(4\pi)^{2d}}\int d\alpha db_1db_2db_3db_4 \frac{\alpha^{2\zeta-1}(b_1b_2b_3b_4)^{\zeta-1}e^{-(\alpha+b_1+b_2+b_3+b_4)}}{\left(\alpha(b_1b_2(b_3+b_4)+b_3b_4(b_1+b_2))+b_1b_2b_3b_4\right)^{\tfrac{d}{2}}}\,.
\end{equation}

However, one needs to take into account the subtraction of the local part of the four-point insertion. Using a Taylor expansion with integral rest, the subtracted $S_{\zeta}$ can then be written as:\footnote{The local part of the four-point insertion responsible for the subdivergence is:
\begin{equation*}
\frac{1}{\Gamma(\zeta)^4\Gamma(2\zeta)(4\pi)^{2d}}\int d\alpha db_1db_2db_3db_4 \frac{\alpha^{2\zeta-1}(b_1b_2b_3b_4)^{\zeta-1}e^{-(\alpha+b_1+b_2+b_3+b_4)}}{\left(\alpha(b_1b_2(b_3+b_4)+b_3b_4(b_1+b_2))\right)^{\tfrac{d}{2}}}\, .
\end{equation*}
Denoting $f(t)=\frac{1}{\Gamma(\zeta)^4\Gamma(2\zeta)(4\pi)^{2d}}\int d\alpha db_1db_2db_3db_4 \frac{\alpha^{2\zeta-1}(b_1b_2b_3b_4)^{\zeta-1}e^{-(\alpha+b_1+b_2+b_3+b_4)}}{\left(\alpha(b_1b_2(b_3+b_4)+b_3b_4(b_1+b_2))+tb_1b_2b_3b_4\right)^{\tfrac{d}{2}}}$, the subtracted integral is thus $f(1)-f(0)$. Using a Taylor expansion with integral rest $f(1)=f(0)+\int_0^1 f'(t) dt$, we obtain \eqref{eq:subtractS} where we still denote the subtracted integral as $S_{\zeta}$.}

\begin{equation}
S_{\zeta}=\frac{-\tfrac{d}{2}}{\Gamma(\zeta)^4\Gamma(2\zeta)(4\pi)^{2d}}\int_0^1 dt\int d\alpha db_1db_2db_3db_4 \frac{\alpha^{2\zeta-1}(b_1b_2b_3b_4)^{\zeta}e^{-(\alpha+b_1+b_2+b_3+b_4)}}{\left(\alpha(b_1b_2(b_3+b_4)+b_3b_4(b_1+b_2))+tb_1b_2b_3b_4\right)^{1+\tfrac{d}{2}}}\, .
\label{eq:subtractS}
\end{equation}

Using two Mellin parameters, the denominator can be written as:

\begin{align}
&\frac{1}{\left(\alpha(b_1b_2(b_3+b_4)+b_3b_4(b_1+b_2))+tb_1b_2b_3b_4\right)^{1+\tfrac{d}{2}}}=\crcr
&\int [dz_1][dz_2] \frac{\Gamma(-z_1)\Gamma(-z_2)\Gamma(z_1+z_2+\tfrac{d}{2}+1)}{\Gamma(1+\tfrac{d}{2})}\frac{(tb_1b_2b_3b_4)^{z_1}(\alpha b_3b_4(b_1+b_2))^{z_2}}{(\alpha b_1b_2(b_3+b_4))^{z_1+z_2+\tfrac{d}{2}+1}}\, .
\end{align}

We can then integrate the Schwinger parameters using formula \eqref{eq:int_class} as well as perform the $t$ integral.

\paragraph{Short-range: $\zeta=1$, $d=3-\epsilon$.}

We obtain:

\begin{align}
S_1=-\frac{3-\epsilon}{2(4\pi)^6\Gamma(\tfrac{5}{2})}&\int_{-\tfrac{1}{2}^-} [dz_1] \int_{-1^-}[dz_2] \Gamma(-z_1)\Gamma(-z_2)\Gamma(z_1+z_2+\tfrac{5-\epsilon}{2})\Gamma(-\tfrac{1}{2}+\tfrac{\epsilon}{2}-z_1)\crcr
& \frac{\Gamma(2+z_1+z_2)^2\Gamma(\tfrac{3}{2}+\tfrac{\epsilon}{2}+z_1+z_2)}{\Gamma(4+2z_1+2z_2)(1+z_1)}\frac{\Gamma(-\tfrac{1}{2}+\tfrac{\epsilon}{2}-z_2)^2\Gamma(-1+\epsilon-z_2)}{\Gamma(-1+\epsilon-2z_2)} \, ,
\end{align}
where we have moved the contours so that all Gamma functions have positive argument.

The poles in $z_1$ and $z_2$ are independent. There is only one pole giving a contribution of order $\epsilon^{-1}$, situated at $z_1=-\tfrac{1}{2}+\tfrac{\epsilon}{2}$ and $z_2=-1+\epsilon$. 
We obtain:

\begin{equation}
S_1=-\frac{2\pi^2}{\epsilon(4\pi)^6}+\mathcal{O}(\epsilon^0) \, .
\end{equation}

\paragraph{Long-range: $\zeta=\tfrac{d+\epsilon}{3}$.}

With the same method, we obtain in the long-range case:

\begin{equation}
S_{d/3}=\frac{1}{2\epsilon(4\pi)^{2d}}\frac{\Gamma(-\tfrac{d}{6})\Gamma(\tfrac{d}{6})^4}{\Gamma(\tfrac{d}{3})^4\Gamma(\tfrac{2d}{3})\Gamma(\tfrac{d}{2})} +\mathcal{O}(\epsilon^0) \, .
\end{equation}

\subsubsection{$I_{\zeta}$ integral}

%
%

%

We now compute the $I$ integral:

\begin{align}
I_{\zeta}=\frac{1}{\Gamma(\zeta)^6(4\pi)^{2d}}&\int da_1da_2db_1db_2 dc_1dc_2 (a_1a_2b_1b_2c_1c_2)^{\zeta-1}e^{-(a_1+a_2+b_1+b_2+c_1+c_2)}\crcr
&\times \frac{1}{\left[c_1c_2(a_1+a_2)(b_1+b_2)+b_1b_2(a_1+a_2)(c_1+c_2)+a_1a_2(b_1+b_2)(c_1+c_2)\right]^{\tfrac{d}{2}}} \, .
\end{align}

The denominator can be written as:

\begin{align}
&\frac{1}{\left[c_1c_2(a_1+a_2)(b_1+b_2)+b_1b_2(a_1+a_2)(c_1+c_2)+a_1a_2(b_1+b_2)(c_1+c_2)\right]^{\tfrac{d}{2}}}= \crcr
&\int \int [dz_1][dz_2] \frac{\Gamma(-z_1)\Gamma(-z_2)\Gamma(z_1+z_2+\tfrac{d}{2})}{\Gamma(\tfrac{d}{2})}\frac{(a_1a_2(b_1+b_2)(c_1+c_2))^{z_1}(b_1b_2(a_1+a_2)(c_1+c_2))^{z_2}}{(c_1c_2(a_1+a_2)(b_1+b_2))^{z_1+z_2+\tfrac{d}{2}}} \, .
\end{align}

\paragraph{Short-range: $\zeta=1$, $d=3-\epsilon$.}

Integrating the Schwinger parameters, we find in the short-range case:

\begin{align}
I_1&=\frac{1}{\Gamma(\tfrac{3}{2})(4\pi)^6}\int_{-\tfrac{1}{2}^-} [dz_1] \int_{-\tfrac{1}{2}^-} [dz_2] \Gamma(-z_1)\Gamma(-z_2)\Gamma(z_1+z_2+\tfrac{3-\epsilon}{2})\crcr 
& \times \frac{\Gamma(\tfrac{\epsilon-1}{2}-z_1-z_2)^2\Gamma(-1+\epsilon-z_1-z_2)}{\Gamma(-1+\epsilon-2z_1-2z_2)}\frac{\Gamma(1+z_2)^2\Gamma(\tfrac{1+\epsilon}{2}+z_2)}{\Gamma(2+2z_2)}\frac{\Gamma(1+z_1)^2\Gamma(\tfrac{1+\epsilon}{2}+z_1)}{\Gamma(2+2z_1)} \, .
\end{align}

The poles in $z_1$ and $z_2$ are not independent. Between, $-\frac{1}{2}^-$ and $\frac{1}{2}^-$, there are three poles in $z_1$: $0\; ,\, -\tfrac{1}{2}+\tfrac{\epsilon}{2}-z_2 \; , \, -1+\epsilon-z_2$. We obtain:

\begin{align}
I_1=&\frac{1}{\Gamma(\tfrac{3}{2})(4\pi)^6}\int_{-\tfrac{1}{2}^-} [dz_2]\Gamma(-z_2)\frac{\Gamma(1+z_2)^2\Gamma(\tfrac{1+\epsilon}{2}+z_2)}{\Gamma(2+2z_2)} \crcr
& \qquad \times \Bigg[ \Gamma(z_2+\tfrac{3-\epsilon}{2})\Gamma(\tfrac{1+\epsilon}{2})\frac{\Gamma(-\tfrac{1}{2}+\tfrac{\epsilon}{2}-z_2)^2\Gamma(-1+\epsilon-z_2)}{\Gamma(-1+\epsilon-2z_2)}\crcr
& \qquad \quad + \Gamma(\tfrac{1}{2}-\tfrac{\epsilon}{2}+z_2)\Gamma(-\tfrac{1}{2}+\tfrac{\epsilon}{2})\frac{\Gamma(\tfrac{1}{2}+\tfrac{\epsilon}{2}-z_2)^2\Gamma(\epsilon-z_2)}{\Gamma(1+\epsilon-2z_2)}\crcr
& \qquad\quad + \Gamma(1-\epsilon+z_2)\Gamma(\tfrac{1}{2}+\tfrac{\epsilon}{2})\frac{\Gamma(\tfrac{1-\epsilon}{2})^2\Gamma(\epsilon-z_2)^2\Gamma(-\tfrac{1}{2}+\tfrac{3\epsilon}{2}-z_2)}{\Gamma(1-\epsilon)\Gamma(2\epsilon-2z_2)} \Bigg]  \, ,
\end{align}
where we have omitted the remaining double integral as it is of order $\mathcal{O}(\epsilon^0)$. 

There is one contribution of order $\mathcal{O}(\epsilon^{-1})$ from the first term from the pole $z_2=-\tfrac{1}{2}+\tfrac{\epsilon}{2}$:
\begin{equation}
\frac{2}{\Gamma(\tfrac{3}{2})(4\pi)^6}\Gamma(-\tfrac{1}{2})\Gamma(\tfrac{1}{2})^4\Gamma(\epsilon)+\mathcal{O}(\epsilon^0) \, .
\end{equation}

For the second term, there are three singular contributions from the poles at $z_2=0$, $z_2=-\tfrac{1}{2}+\tfrac{\epsilon}{2}$ and $z_2=\epsilon$:
\begin{align}
\frac{1}{\Gamma(\tfrac{3}{2})(4\pi)^6}\Bigg[\Gamma(-\tfrac{1}{2})\Gamma(\tfrac{1}{2})^4\Gamma(\epsilon)+\Gamma(\tfrac{1}{2})^4\Gamma(-\tfrac{1}{2})\Gamma(-\epsilon)-2\Gamma(\tfrac{1}{2})^4\Gamma(-\tfrac{1}{2})\Gamma(\epsilon)\Bigg] \, .
\end{align}

For the third term, again three poles give singular contributions, $z_2=0$, $z_2=-\tfrac{1}{2}+\tfrac{3\epsilon}{2}$ and $z_2=\epsilon$:
\begin{align}
\frac{1}{\Gamma(\tfrac{3}{2})(4\pi)^6}\Bigg[\Gamma(-\tfrac{1}{2})\Gamma(\tfrac{1}{2})^4\frac{\Gamma(\epsilon)^2}{\Gamma(2\epsilon)}+2\Gamma(\tfrac{1}{2})^4\Gamma(-\tfrac{1}{2})\Gamma(-\epsilon)+\Gamma(\tfrac{1}{2})^9\Gamma(2\epsilon)\Bigg]\, .
\end{align}

Putting all singular contributions together, we finally obtain:
\begin{equation}
I_1=\frac{1}{(4\pi)^6}\frac{\pi^4}{\epsilon} +\mathcal{O}(\epsilon^0) \, .
\end{equation}

\paragraph{Long-range: $\zeta=\tfrac{d+\epsilon}{3}$.}

Using the same computation method, we find in the long-range case:

\begin{equation}
I_{d/3}=\frac{1}{(4\pi)^{2d}}\frac{\Gamma(\tfrac{d}{6})^9}{2\epsilon\Gamma(\tfrac{d}{3})^9\Gamma(\tfrac{d}{2})} + \mathcal{O}(\epsilon^0) \, .
\end{equation}

\subsubsection{$J_{\zeta}$ integral}

We now compute the $J_{\zeta}$ integral:

\begin{equation}
J_{\zeta}=\frac{1}{\Gamma(\zeta)^6(4\pi)^{2d}}\int da_1da_2da_3db_1db_2dc \frac{(a_1a_2a_3b_1b_2c)^{\zeta-1}e^{-(a_1+a_2+a_3+b_1+b_2+c)}}{\left[a_1a_2a_3(b_1+b_2)+(c(b_1+b_2)+b_1b_2)(a_1a_2+a_1a_3+a_2a_3)\right]^{\tfrac{d}{2}}} \, .
\end{equation}

This integral has a leading divergence in $\mathcal{O}(\epsilon^{-2})$. We thus have to compute the first two singular contributions. 
We write the denominator as:

\begin{align}
&\frac{1}{\left[a_1a_2a_3(b_1+b_2)+b_1b_2(a_1a_2+a_1a_3+a_2a_3)+c(b_1+b_2)(a_1a_2+a_1a_3+a_2a_3)\right]^{\tfrac{d}{2}}} = \crcr
& \int\int \int [dz_1] [dz_2][dz_3] \frac{\Gamma(-z_1)\Gamma(-z_2)\Gamma(-z_3)\Gamma(z_1+z_2+\tfrac{d}{2})\Gamma(z_3+z_2+\tfrac{d}{2})}{\Gamma(\tfrac{d}{2})\Gamma(z_2+\tfrac{d}{2})} \crcr
& \qquad \times \frac{(b_1b_2)^{z_1}(a_1a_2a_3(b_1+b_2)^{z_2}(a_2a_3)^{z_3}}{(c(b_1+b_2))^{z_1+z_2+\tfrac{d}{2}}(a_1(a_2+a_3))^{z_2+z_3+\tfrac{d}{2}}} \, .
\end{align}
We can then integrate the Schwinger parameters using \eqref{eq:int_class}. 

\paragraph{Short-range: $\zeta=1$, $d=3-\epsilon$.}

We obtain:

\begin{align}
J_1&=\frac{1}{(4\pi)^6\Gamma(\tfrac{3-\epsilon}{2})}\int_{-\tfrac{1}{2}^-} [dz_1]\int_{0^-} [dz_2]\int_{-\tfrac{1}{2}^-} [dz_3] \frac{\Gamma(-z_1)\Gamma(-z_2)\Gamma(-z_3)}{\Gamma(z_2+\tfrac{3-\epsilon}{2})}\Gamma(z_1+z_2+\tfrac{3-\epsilon}{2})\Gamma(z_2+z_3+\tfrac{3-\epsilon}{2}) \crcr
& \times \Gamma(\tfrac{-1+\epsilon}{2}-z_1-z_2)\Gamma(\tfrac{-1+\epsilon}{2}-z_3)\frac{\Gamma(1+z_1)^2\Gamma(\tfrac{1+\epsilon}{2}+z_1)}{\Gamma(2+2z_1)}\frac{\Gamma(1+z_2+z_3)^2\Gamma(\tfrac{1+\epsilon}{2}+z_2+z_3)}{\Gamma(2+2z_2+2z_3)} \, .
\end{align}

Between $-\frac{1}{2}^-$ and $\frac{1}{2}^-$, there are two poles in $z_3$: $z_3=0$ and $z_3=-1/2+\epsilon/2$. We thus have:

\begin{align}
J_1&=\frac{1}{(4\pi)^6\Gamma(\tfrac{3-\epsilon}{2})}\int_{-\tfrac{1}{2}^-} [dz_1]\int_{0^-} [dz_2]\Gamma(-z_1)\Gamma(-z_2)\frac{\Gamma(z_1+z_2+\tfrac{3-\epsilon}{2})}{\Gamma(z_2+\tfrac{3-\epsilon}{2})}\Gamma(\tfrac{-1+\epsilon}{2}-z_1-z_2) \crcr
& \times \frac{\Gamma(1+z_1)^2\Gamma(\tfrac{1+\epsilon}{2}+z_1)}{\Gamma(2+2z_1)}\Bigg[ \Gamma(-\tfrac{1}{2}+\tfrac{\epsilon}{2})\Gamma(z_2+\tfrac{3-\epsilon}{2})\frac{\Gamma(1+z_2)^2\Gamma(\tfrac{1+\epsilon}{2}+z_2)}{\Gamma(2+2z_2)} \crcr
& \qquad + \Gamma(\tfrac{1}{2}-\tfrac{\epsilon}{2})\Gamma(1+z_2)\frac{\Gamma(\tfrac{1}{2}+\tfrac{\epsilon}{2}+z_2)^2\Gamma(z_2+\epsilon)}{\Gamma(1+\epsilon+2z_2)}\Bigg] \crcr
&+\frac{1}{(4\pi)^6\Gamma(\tfrac{3-\epsilon}{2})}\int_{-\tfrac{1}{2}^-} [dz_1]\int_{0^-} [dz_2]\int_{\tfrac{1}{2}^-} [dz_3] \frac{\Gamma(-z_1)\Gamma(-z_2)\Gamma(-z_3)}{\Gamma(z_2+\tfrac{3-\epsilon}{2})}\Gamma(z_1+z_2+\tfrac{3-\epsilon}{2})\Gamma(z_2+z_3+\tfrac{3-\epsilon}{2}) \crcr
& \times \Gamma(\tfrac{-1+\epsilon}{2}-z_1-z_2)\Gamma(\tfrac{-1+\epsilon}{2}-z_3)\frac{\Gamma(1+z_1)^2\Gamma(\tfrac{1+\epsilon}{2}+z_1)}{\Gamma(2+2z_1)}\frac{\Gamma(1+z_2+z_3)^2\Gamma(\tfrac{1+\epsilon}{2}+z_2+z_3)}{\Gamma(2+2z_2+2z_3)} \, .
\label{eq:J_1}
\end{align}
With a careful analysis of the poles of the first double integral, we can show that it is of order $\mathcal{O}(\epsilon^0)$ and thus does not contribute to our final result. Likewise, with a long but straightforward computation, we can show that the remaining triple integral is finite in $\epsilon$. 

Let us now compute the second term in \eqref{eq:J_1}. The first poles in $z_2$ are located at $z_2=0$ and $z_2=-\frac{1}{2}+\frac{\epsilon}{2}-z_1$. We obtain: 
\begin{align}
J_1=&\frac{1}{(4\pi)^6\Gamma(\tfrac{3-\epsilon}{2})}\int_{-\tfrac{1}{2}^-} [dz_1]\Gamma(-z_1)\Gamma(\tfrac{1}{2}-\tfrac{\epsilon}{2})\frac{\Gamma(1+z_1)^2\Gamma(\tfrac{1+\epsilon}{2}+z_1)}{\Gamma(2+2z_1)} \crcr
& \times \Bigg[ \frac{\Gamma(\tfrac{1}{2}+\tfrac{\epsilon}{2})^2\Gamma(\epsilon)}{\Gamma(1+\epsilon)\Gamma(\tfrac{3-\epsilon}{2})}\Gamma(z_1+\tfrac{3-\epsilon}{2})\Gamma(\tfrac{\epsilon-1}{2}-z_1) \crcr
& \qquad  +\frac{\Gamma(\tfrac{1-\epsilon}{2}+z_1)}{\Gamma(1-z_1)}\frac{\Gamma(\epsilon-z_1)^2\Gamma(\tfrac{3\epsilon-1}{2}-z_1)\Gamma(\tfrac{1+\epsilon}{2}-z_1)}{\Gamma(2(\epsilon-z_1))} \Bigg] + \mathcal{O}(\epsilon^0)\, ,
\end{align}
where we have omitted the remaining double integral as it is finite in $\epsilon$.
The first term can be written as:
\begin{equation}
\frac{1}{(4\pi)^3\Gamma(\tfrac{3-\epsilon}{2})}\frac{\Gamma(\tfrac{1}{2}-\tfrac{\epsilon}{2})\Gamma(\tfrac{1}{2}+\tfrac{\epsilon}{2})^2}{\Gamma(1+\epsilon)}\Gamma(\epsilon)D_1 \, .
\label{eq:J11}
\end{equation}

The second term has four poles giving singular contributions. The two located at $0$ and $\epsilon$ cancel. The two located at $-\frac{1}{2}+\tfrac{3\epsilon}{2}$ and $-\tfrac{1}{2}+\tfrac{\epsilon}{2}$ give the following contribution:\footnote{There is also a pole at $z_1=\tfrac{1+\epsilon}{2}$ but as it does not lead to a singular contribution, we do not discuss it further.}

\begin{align}
\frac{1}{(4\pi)^6\Gamma(\tfrac{3-\epsilon}{2})}\Bigg[\frac{\Gamma(\tfrac{1}{2}-\tfrac{3\epsilon}{2})\Gamma(\tfrac{1}{2}+\tfrac{3\epsilon}{2})^2\Gamma(\tfrac{1}{2}-\tfrac{\epsilon}{2})^3}{\Gamma(\tfrac{3}{2}-\tfrac{3\epsilon}{2})\Gamma(1+3\epsilon)}\Gamma(\epsilon)\Gamma(2\epsilon) -\frac{\Gamma(\tfrac{1}{2}-\tfrac{\epsilon}{2})^2\Gamma(\tfrac{1}{2}+\tfrac{\epsilon}{2})^4}{\Gamma(\tfrac{3}{2}-\tfrac{\epsilon}{2})\Gamma(1+\epsilon)^2}\Gamma(\epsilon)^2\Bigg] \, .
\label{eq:J12}
\end{align}

Gathering the contributions from \eqref{eq:J11} and \eqref{eq:J12}, we finally obtain for $J_1$: 
%
\begin{equation}
J_1=\frac{2\pi^2}{\left(4\pi\right)^6}\Big[\frac{1}{\epsilon^2}+\frac{2}{\epsilon}\left(2\log(\tfrac{2}{3})+\psi(\tfrac{3}{2})\right)\Big] +\mathcal{O}(\epsilon^0) \, .
\end{equation}

We then obtain for $\alpha_{J_1}$:
\begin{equation}
\alpha_{J_1}= \frac{2\pi^2}{3}\left[ \psi(\tfrac{1}{2})-\psi(\tfrac{3}{2})\right] =-\frac{4\pi^2}{3} \, .
\label{eq:alphaJ1}
\end{equation}

\paragraph{Long-range: $\zeta=\frac{d+\epsilon}{3}$.}

Using the same method, we find for $J_{d/3}$ and $\alpha_{J_{d/3}}$ in the long-range case:
\begin{equation}
J_{d/3}=\frac{\Gamma(\tfrac{d}{6})^6}{2\left(4\pi\right)^{2d}\Gamma(\tfrac{d}{2})^2\Gamma(\tfrac{d}{3})^6}\left[\frac{1}{\epsilon^2}+\frac{1}{\epsilon}\left(3\psi(1)+\psi(\tfrac{d}{6})-5\psi(\tfrac{d}{3})+\psi(\tfrac{d}{2})+K\right)\right] \, ,
\end{equation}
\begin{equation}
\alpha_{J_{d/3}}=\frac{\Gamma(\tfrac{d}{6})^6}{6\Gamma(\tfrac{d}{2})^2\Gamma(\tfrac{d}{3})^6}\Bigg[\psi(\tfrac{d}{6})-\psi(1)+\psi(\tfrac{d}{3})-\psi(\tfrac{d}{2})\Bigg] \, .
\label{eq:alphaJz}
\end{equation}

\section{Conventions for the interaction terms}
\label{ap:conventions}

In this appendix, we write explicitly the interactions appearing in \eqref{eq:invariants}, as well as the quartic invariants, in terms of contraction operators built as linear combinations of products of Kronecker delta functions.

We will use the compact notation $\mathbf{a}=(a_1a_2a_3)$.
The $U(N)^3$ quartic invariants, called pillow and double-trace, respectively, are:
\begin{align}
I_p &=  \delta^p_{\mba\mbb; \mbc\mbd}\phi_{\mba}(x) \phib_{\mbb}(x)  \phi_{\mbc}(x) \phib_{\mbd }(x)\,,\\
I_d &= \delta^d_{\mba\mbb; \mbc\mbd }  \phi_{\mba}(x) \phib_{\mbb}(x)  \phi_{\mbc}(x) \phib_{\mbd }(x)\,,
\end{align}
with: 
\be
    \delta^p_{\mba\mbb; \mbc\mbd }=\frac{1}{3} \sum_{i=1}^3  \delta_{a_id_i} \delta_{b_ic_i} \prod_{j\neq i}  \delta_{a_jb_j}\delta_{c_j d_j} \; , \quad\quad  \delta^d_{\mba\mbb; \mbc\mbd }  = \delta_{\mba \mbb}  \delta_{\mbc \mbd}\,,
\ee
and $\delta_{\mba \mbb}  = \prod_{i=1}^3 \delta_{a_i b_i} $.

The sextic invariants depicted in Fig. \ref{fig:invariants} are:
\begin{align}
I_1 &=  \delta^{(1)}_{\mba\mbb \mbc; \mbd\mbe\mbf} \phi_{\mba}(x) \phib_{\mbd}(x)  \phi_{\mbb}(x) \phib_{\mbe }(x)\phi_{\mbd }(x)\phib_{\mbf }(x)\,,\\
I_b &=\delta^{(b)}_{\mba\mbd; \mbb\mbe; \mbc\mbf }  \phi_{\mba}(x) \phib_{\mbd}(x)  \phi_{\mbb}(x) \phib_{\mbe }(x)\phi_{\mbc }(x)\phib_{\mbf }(x)\,,   \qquad b=2,\ldots,5\,,
\end{align}
with
%
\begin{align}
    \delta^{(1)}_{\mba\mbb\mbc;\mbd \mbe\mbf }=& \d_{a_1d_1}\d_{a_2f_2}\d_{a_3e_3}\d_{b_1e_1}\d_{b_2d_2}\d_{b_3f_3}\d_{c_1f_1}\d_{c_2e_2}\d_{c_3d_3}\, ,\crcr
   \delta^{(2)}_{\mba\mbd; \mbb\mbe; \mbc\mbf }=& \frac{1}{9}\Bigg( \sum_{i=1}^3 \sum_{j \neq i}  \delta_{a_ie_i} \delta_{b_id_i} \delta_{c_j e_j}\delta_{f_jb_j}\left(\prod_{k\neq i}  \delta_{a_kd_k} \right)\left(\prod_{l\neq j}  \delta_{f_l c_l} \right) \left( \prod_{m \neq i,j} \delta_{b_m e_m}\right)  \crcr
  &  \quad  + \mbb\mbe \leftrightarrow \mba\mbd + \mbb\mbe \leftrightarrow \mbc\mbf \Bigg) \, ,\crcr
   \delta^{(3)}_{\mba\mbd; \mbb\mbe; \mbc\mbf }= &\frac{1}{3} \sum_{i=1}^3 \delta_{a_if_i} \delta_{b_id_i} \delta_{c_i e_i}\prod_{j\neq i}  \delta_{a_j d_j}\delta_{b_j e_j}\delta_{c_j f_j} \, ,\crcr
   \delta^{(4)}_{\mba\mbd; \mbb\mbe; \mbc\mbf }=&\frac{1}{3}\left(\delta_{\mba\mbd}\delta^p_{\mbb\mbe;\mbc\mbf} + \delta_{\mbb\mbe}\delta^p_{\mba\mbd;\mbc\mbf}+ \delta_{\mbc\mbf}\delta^p_{\mba\mbd;\mbb\mbe}\right)\, ,\crcr
   \delta^{(5)}_{\mba\mbd; \mbb\mbe; \mbc\mbf }=&\delta_{\mba\mbd}\delta_{\mbb\mbe}\delta_{\mbc\mbf}\, .
 \end{align}
Besides the color symmetrization, to simplify the computation of the beta-functions, we have included a symmetrization with respect to exchange of pairs of black and white vertices.

\section{Stability matrix}
\label{ap:stability}

In this appendix, we give the expression of the full stability matrix $\frac{\partial \beta_i}{\partial \tilde{g}_j}(\tilde{g}^{\star})$ for the $U(N)^3$ short-range model. At order $\tilde{N}^{-1}$ and $\epsilon^{3/2}$ we have:

\begin{equation}
\frac{\partial \beta_1}{\partial \tilde{g}_j}(\tilde{g}^{\star})=\begin{pmatrix}
4\epsilon\\
0 \\
0\\
0\\
0
\end{pmatrix} \, ,
\end{equation}

\begin{equation}
\frac{\partial \beta_2}{\partial \tilde{g}_j}(\tilde{g}^{\star})=\begin{pmatrix}
 \mp 18\sqrt{\epsilon}+\frac{9\sqrt{\epsilon}}{\tilde{N}}\left(\mp3+4\sqrt{\epsilon}\pm20(1+\pi^2)\epsilon\right) \\
 4\epsilon+\frac{4\epsilon}{\tilde{N}}\left(-1\pm2\sqrt{\epsilon}\right)\\
-\frac{12\epsilon}{\tilde{N}}\\
 0\\
 0\\
\end{pmatrix} \, ,
\end{equation}

\begin{equation}
\frac{\partial \beta_3}{\partial \tilde{g}_j}(\tilde{g}^{\star})=\begin{pmatrix}
 \frac{-12\sqrt{\epsilon}}{\tilde{N}}\left(\pm 2+3\sqrt{\epsilon}\right)\\
 \frac{4\epsilon}{\tilde{N}}\\
6\epsilon\\
0 \\
 0\\
\end{pmatrix} \, ,
\end{equation}

\begin{equation}
\frac{\partial \beta_4}{\partial \tilde{g}_j}(\tilde{g}^{\star})=\begin{pmatrix}
\frac{90\sqrt{\epsilon}}{7}\left(\pm2+7\sqrt{\epsilon}\right)+\frac{18\sqrt{\epsilon}}{49\tilde{N}}\left(\pm(103-8\pi^2)-2(203-30\pi^2)\sqrt{\epsilon}\mp (1350 +959\pi^2)\epsilon\right)\\
20\epsilon +\frac{\epsilon}{7\tilde{N}}\left(-60-11\pi^2\mp (112+50\pi^2)\sqrt{\epsilon}\right)\\
24\epsilon+\frac{3\epsilon}{14\tilde{N}}\left(-32-7\pi^2\mp 28\pi^2\sqrt{\epsilon}\right)\\
14\epsilon+\frac{\epsilon}{2\tilde{N}}\left(-16-\pi^2\mp 4\pi^2\sqrt{\epsilon}\right) \\
 0\\
\end{pmatrix} \, , 
\end{equation}

\begin{align}
&\frac{\partial \beta_5}{\partial \tilde{g}_j}(\tilde{g}^{\star})=\crcr
&\begin{pmatrix}
\frac{2\sqrt{\epsilon}}{7}\left(\mp \frac{109}{3}+42\sqrt{\epsilon}\right)\pm \frac{2\sqrt{\epsilon}}{1715\tilde{N}}\left( \frac{1071}{2}+2008\pi^2\pm 294(315-46\pi^2)\sqrt{\epsilon}+ \frac{7}{3}(91656+60655\pi^2)\epsilon\right)\\
6\epsilon+\frac{\epsilon}{49\tilde{N}}\left(\pm 364+45\pi^2+14(28\pm 9\pi^2)\sqrt{\epsilon}\right)\\
\frac{\epsilon}{98\tilde{N}}\left( 504+51\pi^2 \pm 84\pi^2\sqrt{\epsilon} \right) \\
 16\epsilon+\frac{\epsilon}{98\tilde{N}}\left(336+17\pi^2\pm 28\pi^2\sqrt{\epsilon}\right)\\
30\epsilon \\
\end{pmatrix} \, .
\end{align}

The choice of sign is the same as in \eqref{eq:FP_LO}. To obtain the critical exponents, we have to compute the eigenvalues of the above stability matrix. 

\section{Comparison with the sextic $O(N)$ model}
\label{ap:O(N)}

In this appendix, we compare our results with the beta functions of the sextic $O(N)$ model up to four loops. This comparison is not straight-forward. First, we specify the symmetry of the generic sextic multi-scalar model of section \ref{sec:sexticMS} to $U(N)$. Then, we notice that for vector fields $U(N)$ symmetry is equivalent to $O(2N)$ symmetry. We thus substitute $N \rightarrow M/2$ in the $U(N)$ beta functions in order to compare with known results for the $O(N)$ beta functions.

Let us first specify the symmetry in \eqref{eq:beta_SR} to $U(N)$ by setting:
\begin{equation}
\tilde{g}_{\mba \mbb  \mbc ;\mbd \mbe\mbf}=\frac{\tilde{g}}{6}\left(\delta_{ad}\delta_{be}\delta_{cf}+\delta_{ad}\delta_{bf}\delta_{ce}+\delta_{ae}\delta_{bd}\delta_{cf}+\delta_{ae}\delta_{bf}\delta_{cd}+\delta_{af}\delta_{bd}\delta_{ce}+\delta_{af}\delta_{be}\delta_{cd}\right) \; ,
\end{equation}
where each index is now a single index going from $1$ to $N$ and $\mathcal{N}=N$.

We then obtain the following beta function and field critical exponent up to cubic order in $\tilde{g}$:
\begin{equation}
\beta_{U}=-2\epsilon \tilde{g}+\frac{\tilde{g}^2}{48\pi^2}\left(3N+11\right)+\frac{\tilde{g}^3}{9216\pi^4}\left(53N^2-429N-826-\frac{\pi^2}{4}\left(N^3+17N^2+155N+340\right)\right) \, , 
\end{equation}
\begin{equation}
\eta_U=\frac{(N+1)(N+2)\pi^2}{27}\tilde{g}^2 \, .
\end{equation}

We now substitute $N \rightarrow M/2$ and rescale the coupling by $\tilde{g}=\frac{g_O}{5\pi}$ in order to match the conventions of \cite{hager2002six}. We finally obtain:
\begin{equation}
\beta_O=-2\epsilon g_O +\frac{2g_O^2}{15}\left(3M+22\right) -\frac{g_O^3}{1800}\left(8\left(53M^2+858M+3304\right)+\pi^2\left(M^3+34M^2+620M+2720\right)\right) \, ,
\end{equation}
\begin{equation}
\eta_O=\frac{(M+2)(M+4)}{2700}g_O^2 \, ,
\end{equation}
which agrees with the results of \cite{hager2002six}.


\bibliographystyle{JHEP-2.7}

\bibliography{Refs-CFT,Refs-QFT,Refs-TMV}

\addcontentsline{toc}{section}{References}


\end{document}